\documentclass{comnet}

\usepackage{mathrsfs}
\usepackage{blkarray, bigstrut}
\usepackage[demo]{graphicx}

\usepackage{color}

\begin{document}

\title{Evolutionary prisoner's dilemma games coevolving on adaptive networks}


\author{
\name{Hsuan-Wei Lee}
\address{Department of Sociology, University of Nebraska--Lincoln \email{$^*$Corresponding author: waynelee1217@gmail.com}}
\name{Nishant Malik}
\address{Department of Mathematics, Dartmouth College}
\and
\name{Peter J. Mucha}
\address{Department of Mathematics, University of North Carolina at Chapel Hill}}

\maketitle

\begin{abstract}
{We study a model for switching strategies in the Prisoner's Dilemma game on adaptive networks of player pairings that coevolve as players attempt to maximize their return. We use a node-based strategy model wherein each player follows one strategy at a time (cooperate or defect) across all of its neighbors, changing that strategy and possibly changing partners in response to local changes in the network of player pairing and in the strategies used by connected partners. We compare and contrast numerical simulations with existing pair approximation differential equations for describing this system, as well as more accurate equations developed here using the framework of approximate master equations. We explore the parameter space of the model, demonstrating the relatively high accuracy of the approximate master equations for describing the system observations made from simulations. We study two variations of this partner-switching model to investigate the system evolution, predict stationary states, and compare the total utilities and other qualitative differences between these two model variants.}
{Evolutionary games; prisoner's dilemma; adaptive networks; network dynamics; approximate master equations} 
\\
\end{abstract}

\section{Introduction}
Game theory is the study of strategic decision making and the analysis of mathematical models of conflict and cooperation between intelligent rational participants \cite{harsanyi1988general, myerson1991game, gibbons1992primer}. Game theory is used in a variety of disciplines, including biology, economics, political science, computer science, and psychology. In classical game theory, rational actors make their choices to optimize their individual payoffs. That is, actors make strategic choices on a rationally determined evaluation of probable outcomes, or utility, considering the strategic analysis that the players' opponents are making in determining their own choices. Evolutionary game theory \cite{smith1982evolution, nowak1993spatial, weibull1997evolutionary, hofbauer2003evolutionary, hauert2006evolutionary}, the application of game theory to evolving populations, has recently expanded from consideration of lifeforms in biology to various areas in social science. Evolutionary game theory is useful in this context by defining a framework of contests, strategies, behaviors, and analytics in which competition can be modeled. The key point in evolutionary game theory is that the success of a strategy is not just determined by how good the strategy is in itself; rather, it is a question of how good the strategy is in the presence of other alternative strategies and of the distribution of those other strategies within a population.

One of the great difficulties of Darwinian theory, recognized by Darwin himself, was the problem of altruism \cite{mitteldorf2000population}; in particular, exploited cooperators are worse off than defectors. Hence, according to the basic principles of Darwinian selection, intuitively it seems almost certain that cooperation should go extinct. If the basis for selection is at the individual level, the phenomenon of altruism is often hard to interpret. And selection at the group level, or for the ``greater good," appears to violate the game theory assumption of individuals maximizing their own utility. Indeed, altruism is certainly not found to be the general case in nature. Nevertheless, altruistic behaviors can be found in many social animals and, indeed, can be fundamental for some species to survive \cite{fletcher2007evolution}. 

The solution to this apparent paradox can be demonstrated in the application of evolutionary game theory to the prisoner's dilemma game \cite{miller1996coevolution, axelrod2006evolution}, a game which tests the outcomes of ``cooperating" versus ``defecting." Cooperation is usually analyzed in game theory by means of a non-zero-sum game. The prisoner's dilemma game, one of the most studied systems in all of game theory, is a standard example that shows why two completely rational individuals may not cooperate, even if it appears that it is in their best collective interest to do so. Within evolutionary game theory, the analysis of the prisoner's dilemma is as an iterative game \cite{mertens1989repeated}, with the repetitive nature affording competitors the possibility of retaliating or defecting based on the results of previous rounds of the game. There are a multitude of strategies which have been tested by the mathematics of evolutionary game theory and in computer simulations of contests (see, e.g., \cite{fowler2008tournament}), with the general conclusion that the most effective competitive strategies are typically cooperative with a reserved retaliatory response as necessary. The most famous and one of the most successful of these strategies is ``Tit for Tat," which carries out this approach by executing a simple algorithm \cite{axelrod2006evolution,nisan2007algorithmic}.

In recent years, games played on various random graphs and social networks have been investigated \cite{nowak1992evolutionary, killingback1996spatial, abramson2001social, ebel2002coevolutionary, jackson2002formation, hauert2004spatial, lieberman2005evolutionary, santos2006cooperation, fu2008reputation, fu2009partner, hidalgo2015cooperation, pinheiro2016linking}. For example, the public goods game can be seen as a generalization of multiplayer N-player games and group interactions such as pattern formation and self-organization could be studied through this extension \cite{hauert2002replicator, szolnoki2009topology, perc2013evolutionary}. The studies devoted to evolutionary games on complex networks are extensive. Speaking generally, the goals of such studies typically include identifying which combination of game rules, dynamics, and various network topologies can provide cooperation among selfish and unrelated individuals. In the present work, we focus on the dynamical role of players having the option to switch partners, as studied in the framework of adaptive networks that coevolve with the player strategies \cite{pacheco2006active, pacheco2006coevolution, szolnoki2008making, szolnoki2009resolving, szolnoki2009emergence, perc2010coevolutionary}. On a ``coevolving" or ``adaptive" network with a game, the vertices represent players and the edges denote the pairings, or game interactions, between players.

Importantly, the spatial structures in a network could enable cooperators to form small groups or clusters to protect themselves against exploitation by defectors \cite{nowak1992evolutionary}. Studies have shown that the spatial structure can promote cooperation \cite{kerr2002local}, but there is also evidence that suggests that spatial structure may not necessarily favor cooperation \cite{hauert2004spatial}. Other studies have elaborated on different aspects of cooperation on scale-free \cite{gomez2007dynamical, pusch2008impact, perc2009evolution}, square-lattice \cite{perc2008restricted, roca2009promotion}, small-world \cite{kim2002dynamic, santos2005epidemic, szolnoki2009impact}, social and real-world networks \cite{holme2003prisoners, fu2007social, lozano2008mesoscopic}, and multilayer networks \cite{wang2015evolutionary}. Insofar as the player pairing network is part of a larger social setting, other notable rules encouraging cooperative behavior are kin selection \cite{hamilton1964genetical}, group selection \cite{dugatkin1996cooperation,traulsen2004stochastic},  direct reciprocity \cite{pacheco2008repeated}, indirect reciprocity \cite{nowak1998dynamics, nowak2005evolution}, social diversity \cite{perc2008social, santos2008social, yang2009diversity}, voluntary participation \cite{hauert2002volunteering, hauert2007via, chen2008emergence}, reputation \cite{fu2008reputation}, and frequency of breaking cooperator-defector bonds and rewiring \cite{szolnoki2008making, szolnoki2009resolving}. All of these have been studied as interesting mechanisms that may promote cooperation in evolutionary games.

In this paper, we focus our study on a simple coevolving network model for how players play a prisoner's dilemma game with the ability to adapt by changing their strategies or switching partners if they are exploited by their neighbors, introduced in \cite{fu2009partner}. We are interested in exploring the cooperative level among the individuals as they organize into networks of cooperators. By using the technique of approximate master equations, we provide a more accurate approximation of the evolution of the network structure and player states with which to explore the parameter space of the model. We compare the existing analytical methods and our new approximation, providing qualitative and quantitative estimations of various network properties. In Section 2, we introduce the model. We then describe our more accurate approximation in Section 3. In Section 4, we provide numerical results and compare the two approximation methods. In Section 5, we introduce a variation to the model and compare the results of the two variants of the model. Finally, we make concluding remarks in Section 6.

\section{Description of the CD-switching model}

We study the partner switching model introduced by Fu \emph{et al}.~\cite{fu2009partner}, where vertices of the coevolving network represent players, edges denote the pairing of individuals playing the game, and players adapting by changing strategies based on their current local information or switching partners.
For simplicity, we initialize our simulations with an Erd\H{o}s-R\'{e}nyi network with $N$ nodes and $M$ edges, with each node assigned an initial state either as a cooperator ($C$) or defector ($D$). 
Each node uses a single strategy across all of its links, though that strategy can change over time as we describe below. That is, individual $i$ playing with all of her connections obtains an income
$$P_{i} = \sum_{j \in \mathscr{N}_{i}} s_{i}^{T}Ps_{j},$$   
where $\mathscr{N}_{i}$ is the neighborhood set of $i$ and the 2-by-2 payoff matrix $P$ is
\[
\begin{blockarray}{ccc}
& C & D \\
\begin{block}{c(cc)}
  C & 1 & 0\\
  D & 1+u & u \\
\end{block}
\end{blockarray}
 \]
with cost-benefit ratio $u \in (0,1)$ determining the relative outcomes in the payoff matrix. That is, if players $i$ and $j$ in a pairing both play $C$, then both receive the payoff $1$ from this pairing; if player $i$ plays $C$ and player $j$ plays $D$, then player $i$ gets payoff $0$ and player $j$ gets payoff $1+u$; and so on.

\begin{figure} 
\begin{center}
\includegraphics[width=0.45\textwidth]{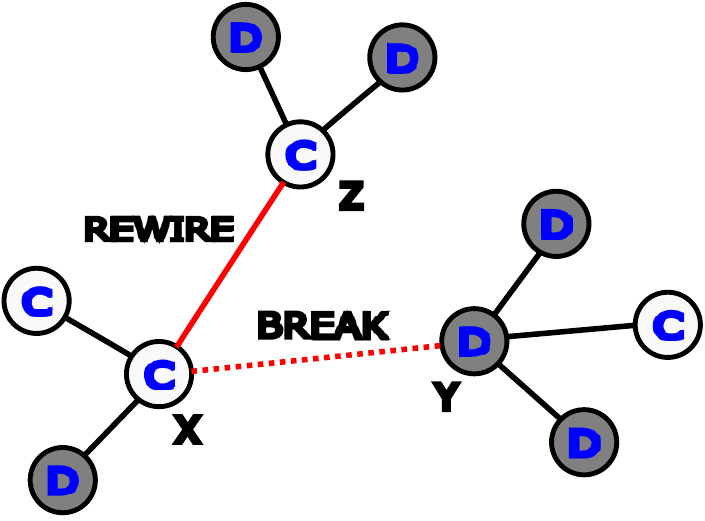}\qquad
\includegraphics[width=0.45\textwidth]{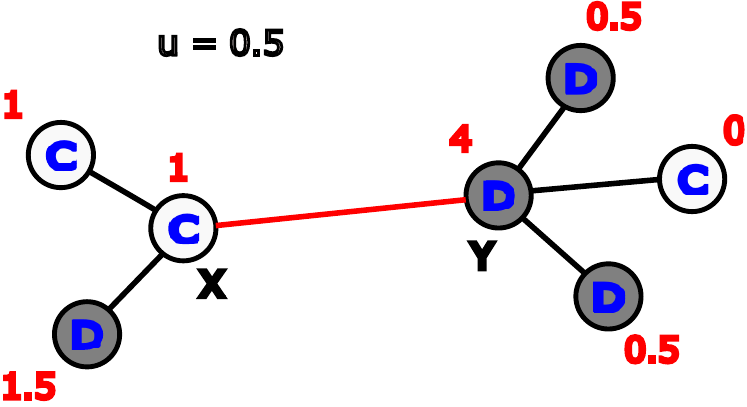}
\caption{Illustrations of the rewiring and strategy updating processes. Pick a discordant edge $XY$ with node $X$ in state $C$ and node $Y$ in state $D$.  (Left) Rewiring occurs with probability $1-w$ with cooperating node $X$ dismissing its defecting neighbor $Y$ and rewiring to a random node $Z$ in the network, independent of the state of node $Z$. If $Z$ is in state $D$, then the number of $CD$ edges remains the same; but if $Z$ is cooperating (as in the picture), then the number of $CD$ edges has been reduced. (Right) Strategy updating occurs with probability $w$. Nodes $X$ and $Y$ compare their utilities, with one node copying the other as selected according to the Fermi function of the difference of utilities. In the example depicted here, suppose the cost-benefit ratio $u = 0.5$ and the Fermi function bias parameter $\alpha=30$. Then node $X$ has utility $P_X=1$, node $Y$ has utility $P_Y=4$, and the Fermi function gives $\phi(s_{X} \leftarrow s_{Y}) = {1}/({1+\exp[\alpha(P_{X}-P_{Y})]}) = {1}/({1+\exp[30(1-4)]}) \approx 1$, so that node $X$ is selected with probability close to 1 to imitate node $Y$'s strategy (state $D$).}
\label{fig:processes}
\end{center}
\end{figure}

Each time step of our microsimulation proceeds as follows. We uniformly at random pick one of the edges that connects a pair of players with different strategies, i.e., a $CD$ link, denoted by $E_{ij}$. (In Section \ref{sec:variation}, we study a variant of the model where both $CD$ and $DD$ links are considered.) With probability $w$ (specified as a parameter), the two nodes $i$ and $j$ connected by edge $E_{ij}$ consider updating their strategies; otherwise (i.e., with probability $1-w$), the edge $E_{ij}$ is rewired. When a node reassesses its strategy along the edge $E_{ij}$, the node has probability $\phi$ specified by the Fermi function to change its state, as first proposed by \cite{szabo1998evolutionary}, as specified in detail below. When link $E_{ij}$ is rewired, the player with end state $C$ unilaterally drops the partnership with its neighbor with end state $D$ on the edge $E_{ij}$ and picks (uniformly at random) another player from the remaining population outside its immediate neighborhood as its new partner. The left panel of Fig.~\ref{fig:processes} illustrates this rewiring process. 

When a strategy updating event occurs, nodes $i$ and $j$ each consider their play across all of their neighbors, observing their total payoffs $P_{i}$ and $P_{j}$. Then the strategy of node $j$ replaces that of $i$ along the edge $E_{ij}$---that is, $i$ copies $j$'s strategy---with probability given by the Fermi function 
 $$\phi(s_{i} \leftarrow s_{j}) = \frac{1}{1+\exp[\alpha(P_{i}-P_{j})]},$$
where $\alpha$ modifies the intensity of the bias towards the strategy with the higher payoff \cite{fudenberg2006evolutionary, traulsen2007stochastic}. Otherwise [that is, with probability $\phi(s_{j} \leftarrow s_{i}) = 1 - \phi(s_{i} \leftarrow s_{j})$], $j$ copies $i$. The value of $1/\alpha$ can be interpreted here as representing the amplitude of noise in the strategy updating process \cite{vukov2006cooperation, ren2007randomness}; that is, $\alpha \rightarrow 0$ ignores the payoffs while $\alpha \rightarrow \infty$ yields deterministic imitation of the node receiving the higher payoff. The right panel of Fig.~\ref{fig:processes} illustrates the strategy updating process.  
 
\begin{figure}
\begin{center}

\includegraphics[trim={0cm 8.5cm 0cm 8.5cm},clip,width=0.75\textwidth]{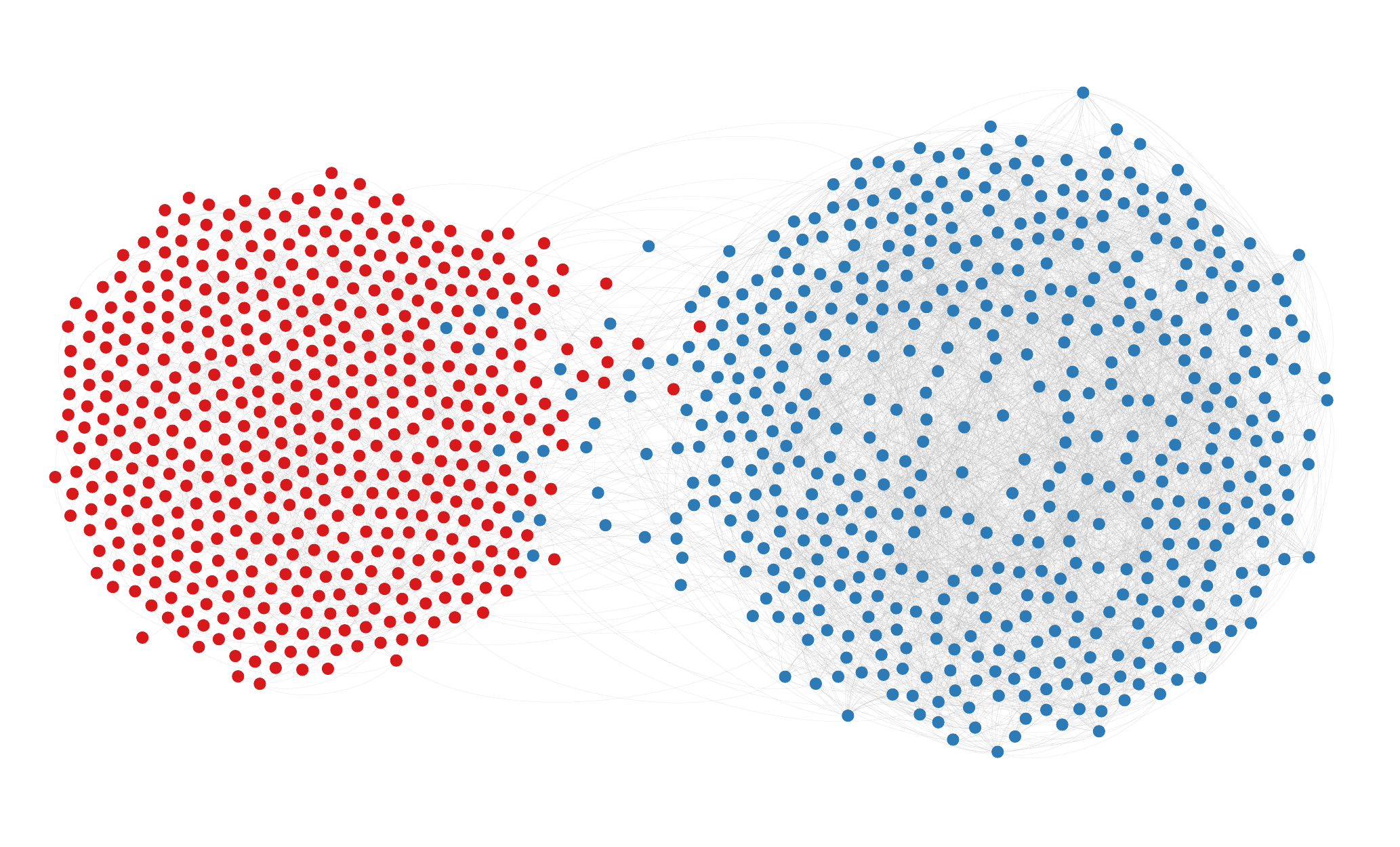}
\caption{Visualization of the CD-switching coevolving network model shortly before the network fissions into two disconnected components, one including all of the remaining cooperator nodes (blue) and the other containing only defecting nodes (red). This simulation contains $N = 1,000$ nodes and $M = 5,000$ edges, with cost-benefit ratio $u = 0.5$, strategy updating probability $w = 0.1$, and initial fraction of defectors $\rho = 0.5$. This visualization was created using the Yifan Hu layout in Gephi \cite{bastian2009gephi}.}
\label{fig:visualization}
\end{center}
\end{figure}

This partner switching evolutionary game stops when there are no discordant edges remaining in the network. That is, only $CC$ and $DD$ edges exist in the final state of the system, with the final network fissioning these two groups (cooperative and defective) into different components. In Fig.~\ref{fig:visualization}, we visualize a simulated network shortly before fission occurs, showing strong grouping of nodes by strategies.

We note that recent work \cite{khoo2016coevolution} studied a spatially-embedded extension of the partner switching model in \cite{fu2009partner}, introducing a range of rewiring distance and finding that a preference for global partner switching can coevolve with cooperation. However, in the present contribution we consider only random rewiring partner switching.

\section{Semi-analytical methods of approximation}
\label{sec:samethods}

We study this generalized coevolving network model with simulations and approximate model equations. The frameworks of Mean Field theory (MF), Pair Approximation equations (PA) and Approximate Master Equations (AME) have all been used effectively in similar settings. The PA equations for these dynamics were obtained by \cite{fu2009partner}. Among these three levels of approximation, AME can often be used to achieve the greatest accuracy \cite{gleeson2013binary}, motivating us to develop the AME system here.

For comparison with the AME system that we will specify, we first look at the PA equations derived in \cite{fu2009partner}, describing the dynamics of the quantities $N_{X}$ and $N_{XY}$---where $X,Y \in \{ C, D\}$---that count the numbers of nodes and edges corresponding to the two node states. For example, $N_{C}$ denotes the number of cooperators and $N_{CC}$ denotes the number of $CC$ links in the network. While the node states and network topology coevolve, the total numbers of nodes ($N$) and edges ($M$) remain conserved by the specified dynamics, requiring $N_{C} + N_{D} = N$ and $N_{CC} + N_{CD} + N_{DD} = M$. The evolution of these counts is then obtained through a moment closure approximating triple counts $N_{XYZ}$---where $X, Y, Z \in \{ C, D\}$---in terms of the edge counts $N_{CC}$, $N_{CD}$ and $N_{DD}$. Specifically, $N_{XXY}$ is approximated by the product of $XX$ links and the average number of $XY$ links that an $X$ node has, i.e. $N_{XY}/N_{X}$, hence $N_{XXY} = 2N_{XX}N_{XY}/N_{X}$. The resulting PA equations \cite{fu2009partner} are as follows:
\begin{equation}
\begin{split}
\frac{dN_{C}}{dt} & = w\cdot N_{CD}\cdot \tanh \bigg[ \frac{\alpha}{2}( \overline{\pi}_{C} - \overline{\pi}_{D}) \bigg] \\
\frac{dN_{CC}}{dt} & = w\cdot \bigg( N_{CD}\phi_{C \rightarrow D} - 2N_{CD}\frac{N_{CC}}{N_{C}}\phi_{D \rightarrow C} + N_{CD}\frac{N_{CD}}{N_{D}}\phi_{C \rightarrow D} \bigg) \\
& + (1-w) \cdot \frac{N_{C}}{N}N_{CD}  \\
\frac{dN_{DD}}{dt} & = w \cdot \bigg( N_{CD}\phi_{D \rightarrow C} - 2N_{CD}\frac{N_{DD}}{N_{D}}\phi_{C \rightarrow D} + N_{CD}\frac{N_{CD}}{N_{C}}\phi_{D \rightarrow C} \bigg)\,, 
\end{split}
\label{eq:PA}
\end{equation}
where $w$ is the probability of strategy updating (versus edge rewiring), $\overline{\pi}_{C} = 1 \cdot {2N_{CC}}/{N_{C}} + 0 \cdot {N_{CD}}/{N_{D}} $ is the average utility of $C$ nodes, $\overline{\pi}_{D} = (1+u) \cdot {N_{CD}}/{N_{D}} + u \cdot {2N_{DD}}/{N_{D}} $ is the average utility of $D$ nodes, 
$\phi_{C\to D} = 1/(1+\exp [\alpha (\overline{\pi}_{D} - \overline{\pi}_{C})])$, $\phi_{D\to C} = 1/(1+\exp [\alpha (\overline{\pi}_{C} - \overline{\pi}_{D})])$, and $\alpha$ controls the intensity of selection.  

The AME method has been successfully used to approximate various processes on static and coevolving networks. The difficulty in deriving the AME system for the dynamics studied here is that the present transition probabilities depend on more than nearest neighbor states. For comparison, in the SIS model the recovery rate of an infected (I) node and the probability of infection along a given SI edge are both specified as constant parameters. Similarly, in typical voter model dynamics, the probability for a node to change its state relies on its neighbors' opinions. In contrast, in the partner switching evolutionary game studied here, the local information impacting the decision of a node to change its state includes its total utility and that of its neighbors, which depend on the states of the neighbors' neighbors. Since the binary-state AME framework only includes the node states, degrees, and neighbor states, we need estimate neighbors' utilities in terms of these limited quantities. We will explain how we approximate these utilities after we introduce the equations below. 

Let $C_{k,l}(t)$ and $D_{k,l}(t)$, respectively, be the number of cooperating and defecting nodes of degree $k$ with $l$ defecting neighbors at time $t$. We note that the numbers of nodes using each strategy are then given by the zeroth moments of the $C_{k,l}(t)$ and $D_{k,l}(t)$ distributions, $N_{C} = \sum_{kl}C_{k,l}$ and $N_{D} = \sum_{kl}D_{k,l}$. Meanwhile, the first moments give the numbers of edges of each type: $N_{CC} = \frac{1}{2} \sum_{kl}(k-l)C_{k,l}$, $N_{CD} = \sum_{kl}lC_{k,l} = \sum_{kl}(k-l)D_{k,l}$ and $N_{DD} = \frac{1}{2} \sum_{kl}lD_{k,l}$. Importantly, while the node states and network topology coevolve, the total numbers of nodes and edges remain conserved by the specified dynamics, requiring $N_{C} + N_{D} = N$ and $N_{CC} + N_{CD} + N_{DD} = M$. 

\begin{figure}
\begin{center}
\includegraphics[width=0.6\textwidth]{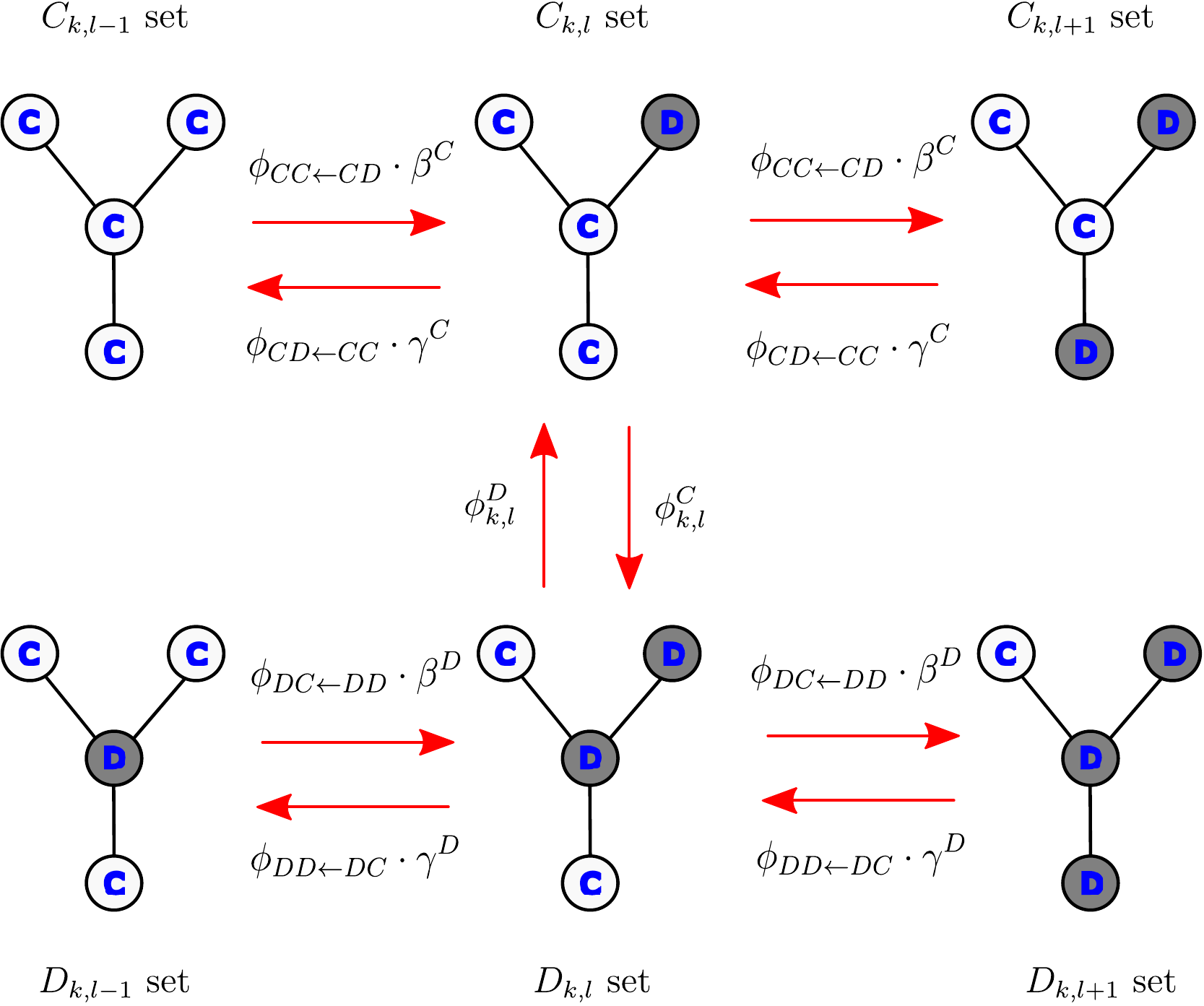}
\caption{Illustration of transitions to and from the $C_{k,l}$ and $D_{k,l}$ compartments (cooperating and defecting nodes, respectively, with degree $k$ and $l$ defecting neighbors) in the system of Approximate Master Equations due to players updating their strategies. For each compartment, only a subset of neighbors in the corresponding ego network are depicted here, to signal the essential changes corresponding to each flow. The term describing each flow is indicated above/below the corresponding red line connecting compartments.}
\end{center}
\label{fig:AME}
\end{figure}

The AME system of ordinary differential equations governing the $C_{k,l}(t)$ and $D_{k,l}(t)$ compartments is of course more complicated than the (relatively) compact PA equations \eqref{eq:PA}. In the following paragraphs, we describe each of the terms that appear in this AME system:
\begin{equation}
\begin{split}
\frac{dC_{k,l}}{dt} & = w \bigg\{ \phi^{D}_{k,l} \cdot (k-l)D_{k,l} - \phi^{C}_{k,l} \cdot l C_{k,l}\\
& + \phi_{CD \leftarrow CC} \cdot \gamma^{C}(l+1) C_{k,l+1} - \phi_{CD \leftarrow CC} \cdot \gamma^{C} l C_{k,l}\\
& + \phi_{CC \leftarrow CD} \cdot \beta^{C}(k-l+1) C_{k,l-1} - \phi_{CC \leftarrow CD} \cdot \beta^{C}(k-l) C_{k,l} \bigg \}\\
& + (1-w) \bigg \{  \frac{N_{C}}{N}  \big[(l+1) C_{k,l+1} - l C_{k,l}\big]
 + \frac{N_{CD}}{N_{}} \big[C_{k-1,l} - C_{k,l}\big] \bigg \}
\,,
\end{split}
\label{eq:AME-C}
\end{equation}
\begin{equation}
\begin{split}
\frac{dD_{k,l}}{dt} & = w \bigg\{ -\phi^{D}_{k,l} \cdot (k-l) D_{k,l} + \phi^{C}_{k,l} \cdot l C_{k,l}\\
& + \phi_{DD \leftarrow DC} \cdot \gamma^{D}(l+1) D_{k,l+1} - \phi_{DD \leftarrow DC} \cdot \gamma^{D} l D_{k,l}\\
& + \phi_{DC \leftarrow DD} \cdot \beta^{D}(k-l+1) D_{k,l-1} - \phi_{DC \leftarrow DD} \cdot \beta^{D} (k-l) D_{k,l} \bigg \}\\
& + (1-w) \bigg\{ \big[(k-l+1) D_{k+1,l} - (k-l) D_{k,l}\big]
 + \frac{N_{CD}}{N_{}} \big[D_{k-1,l} - D_{k,l}\big] \bigg \}
\,.
\end{split}
\label{eq:AME-D}
\end{equation}
To describe the AME derivation, we focus on explaining the $C_{k,l}$ equation, as the effects of corresponding terms of the $D_{k,l}$ equation are similar. The first three lines of equation (\ref{eq:AME-C}), pre-multiplied by $w$, are the effect of strategy updating, while the last line with $1-w$ are the effect of rewiring. To write the AME system, we track all counts flowing in and out of ``center" class, $C_{k,l}$. As visualized in Fig.~\ref{fig:AME}, there are six such flows for the $C_{k,l}$ compartment, yielding the first six terms in the $C_{k,l}$ equation. The first term describes the rate at which $D_{k,l}$ nodes change to $C_{k,l}$ through comparing its utility with one of its $C$ neighbors, with the Fermi function evaluation $\phi^{D}_{k,l}$ as estimated below. Similarly, the second term captures $C_{k,l}$ nodes changing strategy to become $D_{k,l}$ after comparing their utility with their $D$ neighbors, with Fermi function evaluation $\phi^{C}_{k,l}$ also estimated below. 

To estimate these Fermi function transition probabilities, we can directly compute the utility of the center node, since we know the numbers and types of its neighbors; but we need to estimate neighbors' utilities in terms of the limited information available in the AME framework. Recall that the payoff matrix of our Prisoner's dilemma game is
\[
\begin{blockarray}{ccc}
& C & D \\
\begin{block}{c(cc)}
  C & 1 & 0\\
  D & 1+u & u \\
\end{block}
\end{blockarray}
 \]
where $u \in (0,1)$. For transitions by a strategy update out of the $D_{k,l}$ class, the center node is of state $D$, with $k$ neighbors, $l$ of which are of state $D$. Hence we can compute the center's utility as 
\[ 
P_{D} = l \cdot u + (k - l) \cdot (1+u). 
\]
Since we do not have any information about the neighbors of the $C$ neighbors of these $D_{k,l}$ nodes, we approximate utility by estimating the numbers and types of the neighbors of a $C$ neighbor. Specifically, we denote the expected number of $C$ neighbors of this $C$ neighbor by $\eta^{D}$, and the expected number of $D$ neighbors of a $C$ neighbor by $\beta^{D} + 1$, where $\eta^{D}$ and $\beta^{D}$ will be estimated below and the $+1$ accounts for the $D_{k,l}$ node itself. The estimate of the utility of the $C$ neighbor thus becomes
\[
\hat P_{C} = \eta^{D} \cdot 1 + (\beta^{D} + 1) \cdot 0,
\]
and we compare the center node $D_{k,l}$'s utility with the estimate from one of its $C$ neighbors via the Fermi function, so that the estimated probability of the $D_{k,l}$ node changing its state in this consideration is 
\[
\phi^{D}_{k,l} = \frac{1}{1+\exp[\alpha(P_{D}-\hat P_{C})]} = 
\frac{1}{1+\exp[\alpha(l u + (k - l) (1+u) - \eta^{D})]}.
\]

Meanwhile, transitions by a strategy update out of the $C_{k,l}$ rely on the utility of a $C_{k,l}$ node,
\[
P_{C} = l \cdot 0 + (k - l) \cdot 1\,,
\]
and estimating the utility of one of the $D$ neighbors via the numbers and types of its neighbors. We denote the expected number of $C$ neighbors of this $D$ neighbor (including the $C_{k,l}$ node in question) by $\gamma^{C}$ + 1, and the number of $D$ neighbors of this $D$ neighbor as $\delta^{C}$, both of which will be estimated below. With this notation, the expected utility of one of the $D$ neighbors is 
\[
\hat P_{D} = (\gamma^{C} + 1) \cdot (1+u) + \delta^{C} \cdot u\,,
\]
and the Fermi function estimate for the probability of the $C_{k,l}$ node changing its state in this comparison becomes
\[
\phi^{C}_{k,l} = \frac{1}{1+\exp[\alpha(P_{C}-\hat P_{D})]} = 
\frac{1}{1+\exp[\alpha( k - l - (\gamma^{C} + 1) (1+u) + \delta^{C} u)]}\,.
\]

The third to sixth terms in the $C_{k,l}$ equation are the effects of one of the neighbors changing its state and thus leading to a change of the total quantity of $C_{k,l}$. Similar to the above, in order to make estimates of the transition probabilities we need a variety of estimates of the numbers and types of the neighbors' neighbors, which we denote as follows. 
We use $\beta^{C}$ to provide the expected number of $D$ neighbors of a $CC$ edge, as 
$$\beta^{C} = \frac{\sum_{k,l}(k-l)lC_{k,l}}{\sum_{k,l}(k-l)C_{k,l}}\,.$$
Similarly, $\beta^{D}$ denotes the expected number of $D$ neighbors of the $C$ node in a $CD$ edge by
$$\beta^{D} = \frac{\sum_{k,l}l^{2}C_{k,l}}{\sum_{k,l}lC_{k,l}}\,.$$
Meanwhile, $\gamma^{C}$ and $\gamma^{D}$ estimate the numbers of $C$ neighbors of the $D$ node of a $CD$ edge or of a $DD$ edge, respectively:
$$\gamma^{C} = \frac{\sum_{k,l}(k-l)^{2}D_{k,l}}{\sum_{k,l}(k-l)D_{k,l}}\,,$$
$$\gamma^{D} = \frac{\sum_{k,l}l (k-l) D_{k,l}}{\sum_{k,l}lD_{k,l}}\,.$$
Similarly, $\delta^{C}$ and $\delta^{D}$ provide the expected numbers of $D$ neighbors of the $D$ node of a $CD$ edge or of a $DD$ edge, respectively, while $\eta^{C}$ and $\eta^{D}$ give the numbers of $C$ neighbors of a $CC$ edge or the $C$ node of a $CD$ edge, respectively:
$$\delta^{C} = \frac{\sum_{k,l}(k-l)lD_{k,l}}{\sum_{k,l}(k-l)D_{k,l}}\,,$$
$$\delta^{D} = \frac{\sum_{k,l}l^{2}D_{k,l}}{\sum_{k,l}lD_{k,l}}\,,$$
$$\eta^{C} = \frac{\sum_{k,l}(k-l)^{2}C_{k,l}}{\sum_{k,l}(k-l)C_{k,l}}\,,$$
$$\eta^{D} = \frac{\sum_{k,l}l (k-l) C_{k,l}}{\sum_{k,l}lC_{k,l}}\,.$$

To compute the corresponding $\phi$ transition probability in each term, let us first consider the Fermi function factors on the second line in equation (\ref{eq:AME-C}) specified by
$$\phi_{CD \leftarrow CC} = \frac{1}{1 + \exp [\alpha (P^{CD \leftarrow CC}_{D} - \pi_{C})]},$$
where 
$$P^{CD \leftarrow CC}_{D} = (\gamma^{C} + 1) \cdot (1+u) + \delta^{C} \cdot u$$
and 
$$\pi_{C} = 1 \cdot \frac{2N_{CC}}{N_{C}} + 0 \cdot \frac{N_{CD}}{N_{C}}.$$
That is, this transition probability considers a $D$ neighbor of the center $C_{k,l}$ class, and we suppose this $D$ neighbor has $\gamma^{C}+1$ neighbors in state $C$ (including the center $C_{k,l}$ node) and $\delta^{C}$ in state $D$. Then we estimate this $D$ node's utility and compare it with its typical $C$ neighbor (that is, the average utility $\pi_{C}$). 

We note for emphasis here that other choices could have been made in this estimate; specifically, we explicitly know that one of the $C$ neighbors of this $D$ node is the center $C_{k,l}$ node. An alternative formulation of the AME system could be formed that separates out the state change of the $D$ neighbor in comparison with the $C_{k,l}$ center (thus removing the +1 contribution) from its state changes through interaction with other $C$ nodes. In order to keep the equations just a little simpler, and because of the good approximation we observe in our results below, we continue to employ the simpler estimate here.

We similarly estimate the Fermi function factors on the third line in equation (\ref{eq:AME-C}) as
$$\phi_{CC \leftarrow CD} = \frac{1}{1 + \exp [\alpha (P^{CC \leftarrow CD}_{C} - \pi_{D})]},$$
where 
$$P^{CC \leftarrow CD}_{C} = (\eta^{C} + 1) \cdot 1 + \beta^{C} \cdot 0$$
and 
$$\pi_{D} = (1+u) \cdot \frac{N_{CD}}{N_{D}} + u \cdot \frac{2N_{DD}}{N_{D}}.$$ 
In analogy to the above argument, we consider a $C$ neighbor of the center $C_{k,l}$ class and suppose this $C$ node has $\eta^{C}+1$ neighbors in state $C$ (including the center node) and $\beta^{C}$ in state $D$. We estimate this $C$ node's utility and compare it with the average utility of a $D$ node, $\pi_{D}$. 

\begin{figure}
\begin{center}
\includegraphics[width=0.45\textwidth]{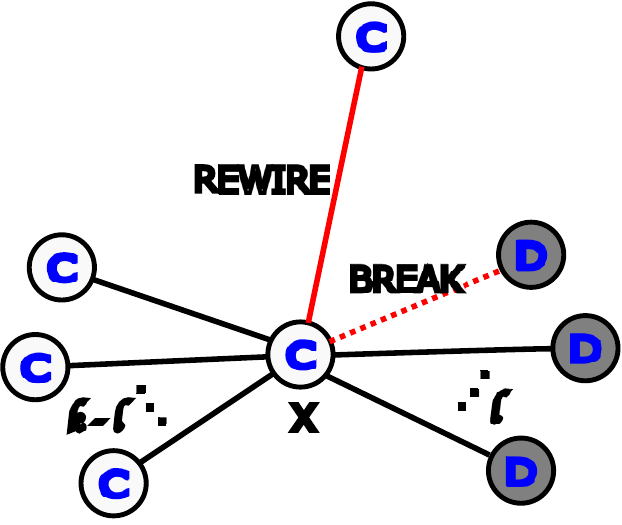}\qquad
\includegraphics[width=0.45\textwidth]{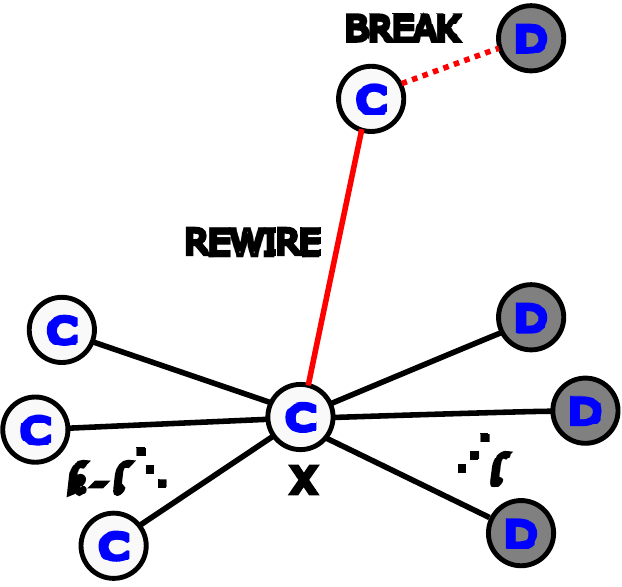}
\caption{Illustrations of nodes undergoing active and passive rewiring. (Left) Active rewiring: before the rewiring, node $X$ is in class $C_{k,l}$ (cooperator of degree $k$ with $l$ defecting neighbors). Suppose one of $X$'s discordant edges is selected and X actively breaks that edge shared with a defective neighbor, rewiring to a random node in the network. If node $X$ rewires to a node of state $C$ (as depicted here), then node $X$ moves to class $C_{k,l-1}$. We recall that only $C$ nodes actively rewire in the present model. (Right) Passive rewiring: before the rewiring, node $X$ is in class $C_{k,l}$. Suppose in the rewiring process, node $X$ is passively rewired through the activity of another $C$ node in the network. Then node $X$ moves to class $C_{k+1,l}$.}
\end{center}
\label{fig:rewiring}
\end{figure}

The last line in the $C_{k,l}$ equation (\ref{eq:AME-C}) describes the rewiring effect. The class $C_{k,l+1}$ flows to $C_{k,l}$ when the center node $C$ drops one of its defecting neighbors and rewires randomly to another $C$. Similarly, the $C_{k,l}$ count decreases when the center node $C$ drops one of its defecting neighbors and rewires randomly to another $C$. If the rewiring instead links to another $D$ node, these counts do not change. Finally, a $C_{k-1,l}$ node becomes $C_{k,l}$ if it is the recipient of a rewired edge (from another $C$ node), while the $C_{k,l}$ count decreases if a $C_{k,l}$ receives a rewired edge. The first group of square-bracketed terms on this line corresponds to the center node actively rewiring to some other node in the network while the second group in square brackets describes the center node passively receiving a rewired edge through the action of some other node in the network. Fig.~\ref{fig:rewiring} illustrates these active and passive rewiring processes. 

Shifting our focus to the six flows in and out of the set depicted in the center of the lower row of Fig.~\ref{fig:AME}, we similarly obtain equation \eqref{eq:AME-D} for the $D_{k,l}$ compartment. The resulting system contains $2(k_\mathrm{max} + 1)^2$ coupled differential equations, where $k_\mathrm{max}$ is the maximum degree a node can have in the network. Usually, the necessary value of $k_\mathrm{max}$ for good accuracy depends on network structure and the mean degree. Here we choose $k_\mathrm{max} = 50$. We numerically solve these differential equations to semi-analytically approximate the evolution of the system, using the ode45 solver in MATLAB until the solutions reach steady state. 

While we consider PA and AME approaches here, we note that \cite{zhou2013link} used an approach for studying adaptive SIS networks that treated the links as the objects, classifying them according to the disease states while tracking the degree and number of infected neighbors at both end nodes. This link-based method could generally improve the accuracy compared to AME; however, it also increases the number of coupled equations from $O(k^{2}_\mathrm{max})$ to $O(k^{4}_\mathrm{max})$, where $k_\mathrm{max}$ is the maximum degree in the network.

\section{Simulations of the CD-switching model}

To test the accuracy of the approximations, we study simulated dynamics on networks with $N = 1,000$ nodes and $M = 5,000$ edges. That is, the mean degree of the network is fixed to be $\langle k \rangle = 2M/N = 10$. We set the parameter $\alpha=30$ in the Fermi function used for imitating strategies. We study dynamics for different initial fractions of defectors, denoted by $\rho$. 
We consider $\rho = 0.5$ first, and explore the influence of this parameter later. We draw the initial networks from an Erd\H{o}s-R\'{e}nyi $G(N, M)$ random graph model of $N$ nodes and $M$ edges distributed uniformly and independently between the nodes. For large $N$, the initial degree distribution is approximated by the Poisson distribution of mean $\langle k \rangle$, with the probability of a selected node having degree $k$ being
$p_{k} = {\langle k \rangle^{k}e^{-\langle k \rangle}}/{k!}.$
In the present model, updating only discordant edges (i.e., $CD$), the dynamics stop when there are no discordant edges remaining.
Unless stated otherwise, we perform $1,000$ simulations for each condition and report the average.  

\subsection{Final level of cooperation}

\begin{figure}
\begin{center}
\includegraphics[width=0.475\textwidth]{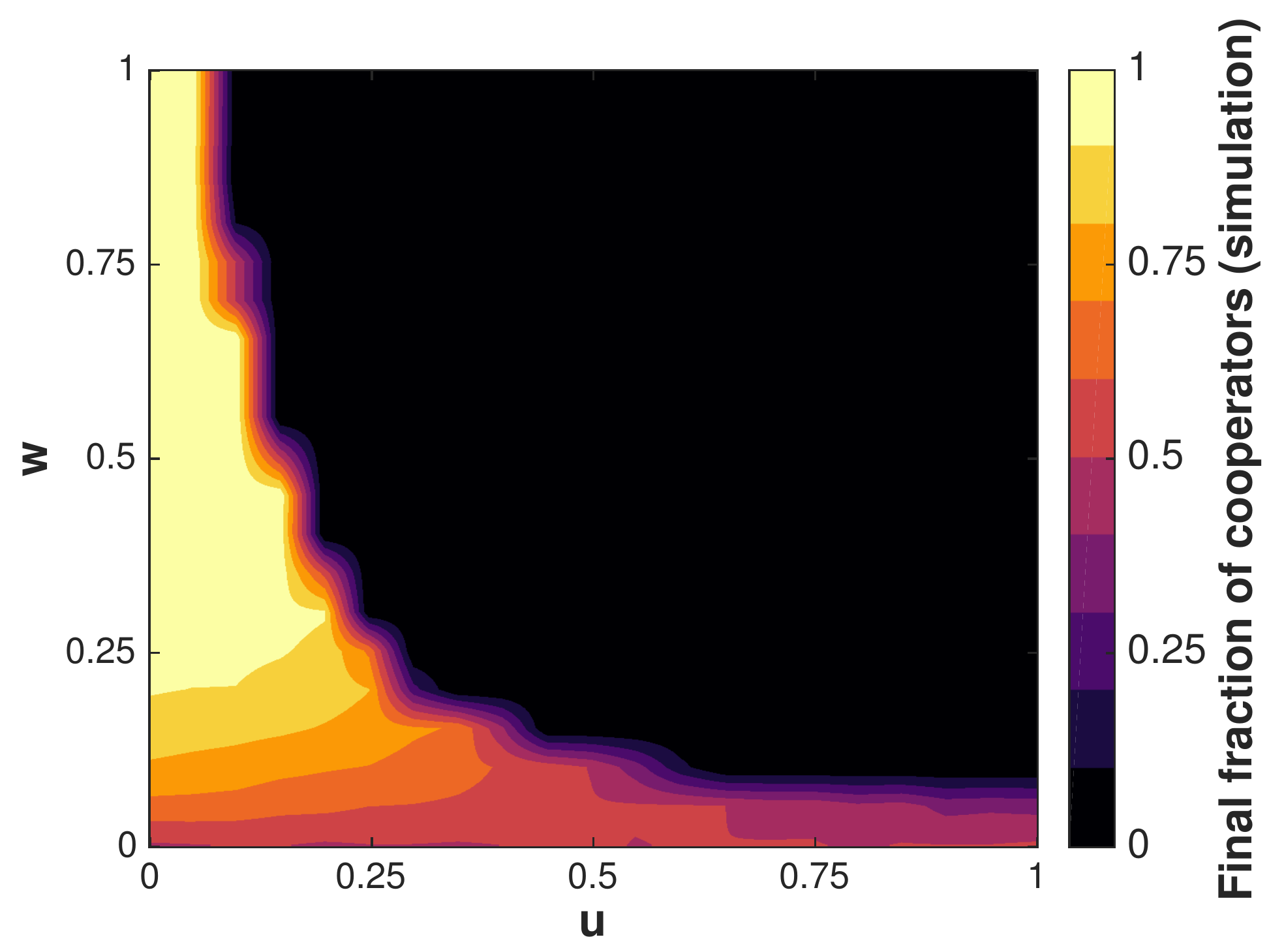}\quad
\includegraphics[width=0.475\textwidth]{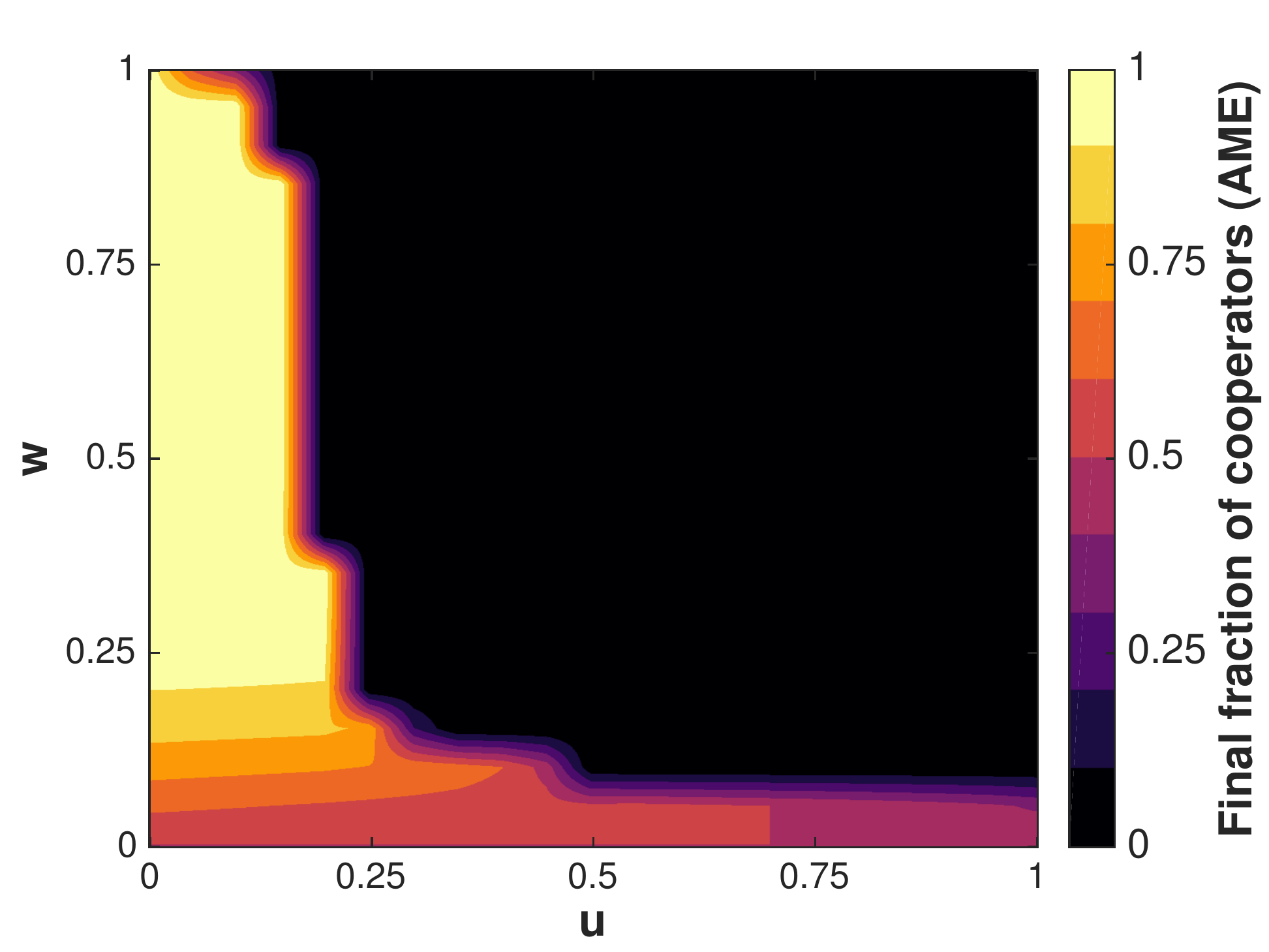}

\includegraphics[width=0.475\textwidth]{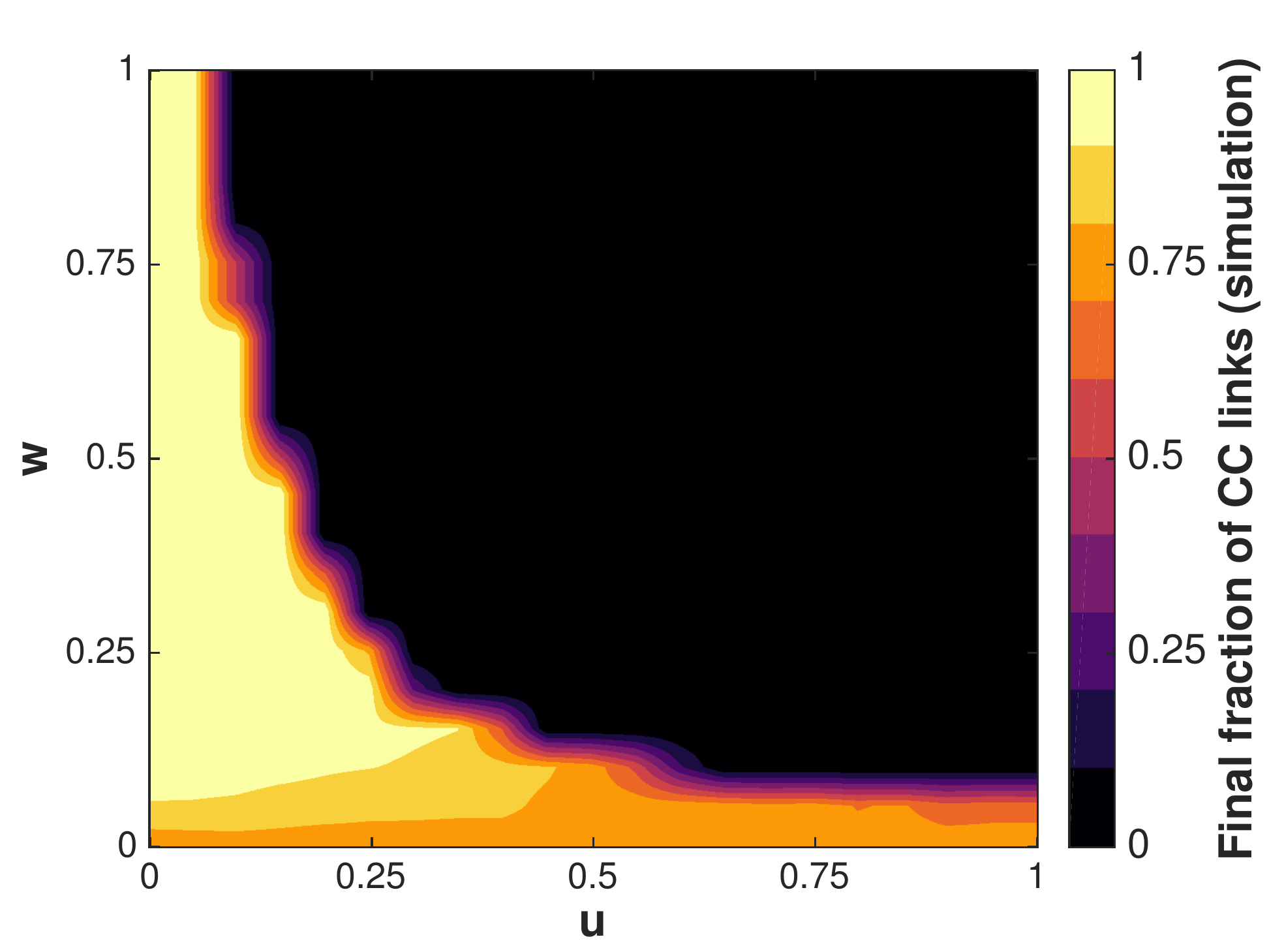}\quad
\includegraphics[width=0.475\textwidth]{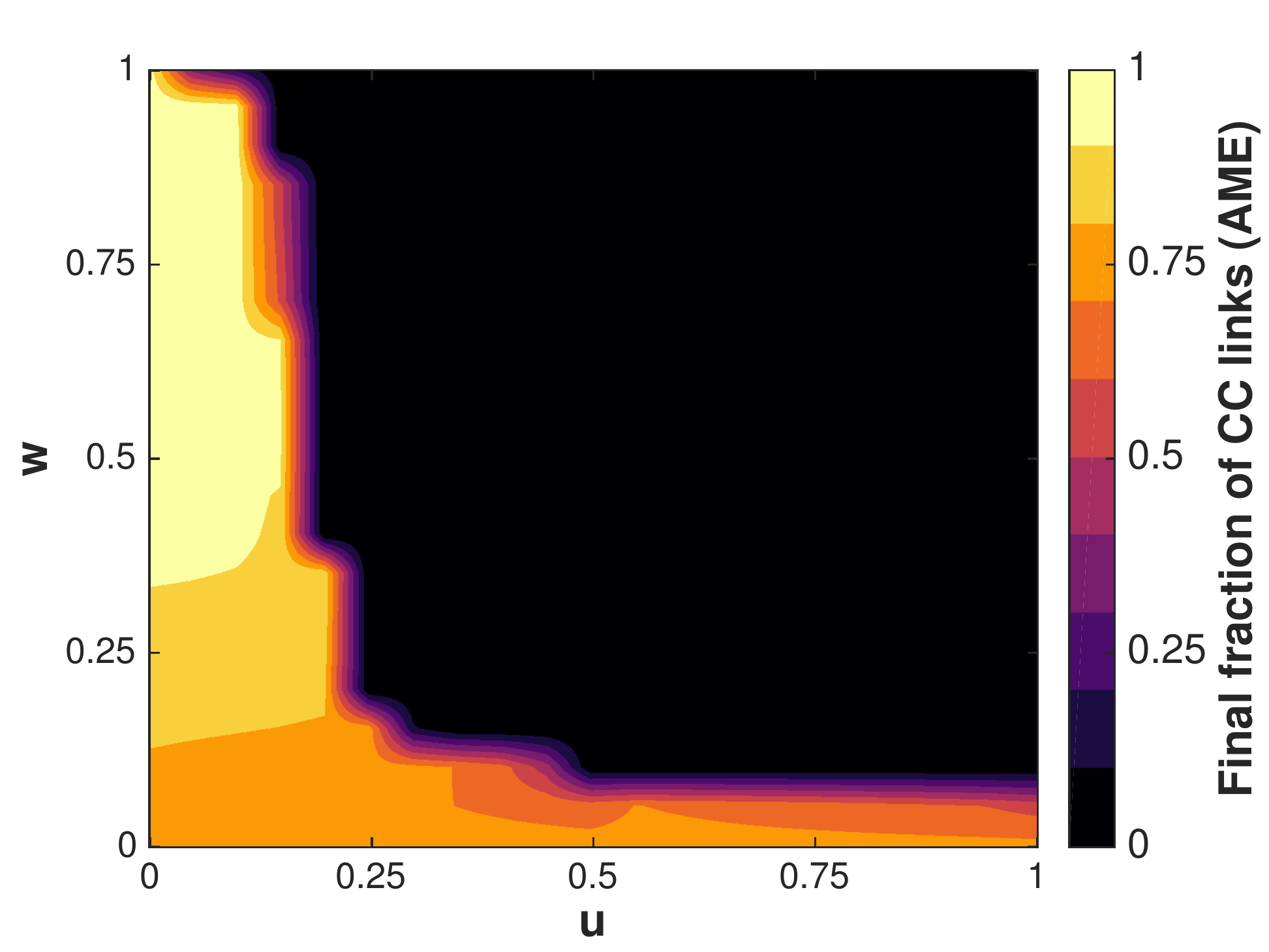}
\caption{Results from simulations (left column) and Approximate Master Equations (AME) (right column) of the final fractions of cooperators (top row) and CC edges (bottom row) for different combinations of the cost-benefit ratio $u$ and strategy updating probability $w$. The initial fraction of defectors is $\rho=0.5$. For both the $u$ and $w$ axes, we use steps of 0.05 and plot the results from stationary states. Simulation results here are averaged over 50 realizations at each parameter set. These visualizations were generated from results on a regular grid through bilinear interpolation, leading to some clearly apparent grid artifacts. While some discrepancies between simulation and AME results are clearly present, we note in particular that the position of the phase transition in the $(u,w)$ parameter space is well approximated by the AME system.}
\label{fig:phasediagram}
\end{center}
\end{figure}

To study the effect of the cost-benefit ratio $u$ and strategy update rate $w$, we focus on the level of cooperation obtained in the final (stationary) states of the model, starting from an initial fraction of defectors $\rho=0.5$. In Fig.~\ref{fig:phasediagram}, we present heat maps generated by simulations and AME approximations of the final fractions of cooperating nodes and CC links. As qualitatively expected, the final fraction of cooperating nodes is higher when the possible additional payoff for defecting, $u$, is smaller. When $w$ is close to $0$, we observe a moderate level of cooperators, as expected because the discordant edges can resolve themselves by rewiring with fewer strategy updates. 
Away from the small values of $u$ or $w$, the final network states are dominated by defectors. The AME approximation generally captures these high and low cooperation regions of the $(u,w)$ parameter space, up to a modest displacement in the precise position of the phase transition.

\begin{figure}
\begin{center}
\hspace*{-0.1in}
\includegraphics[width=0.5\textwidth]{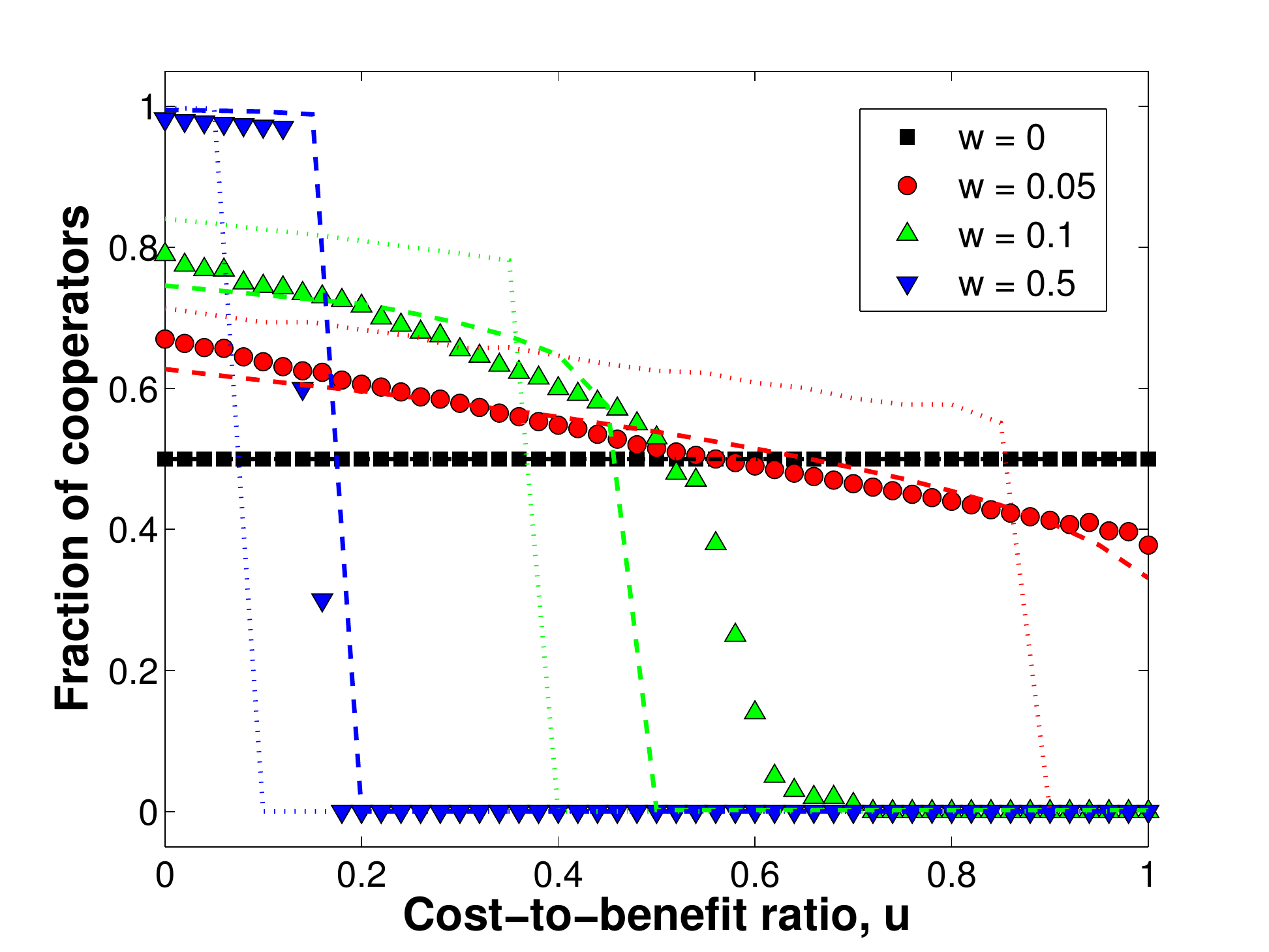}
\includegraphics[width=0.5\textwidth]{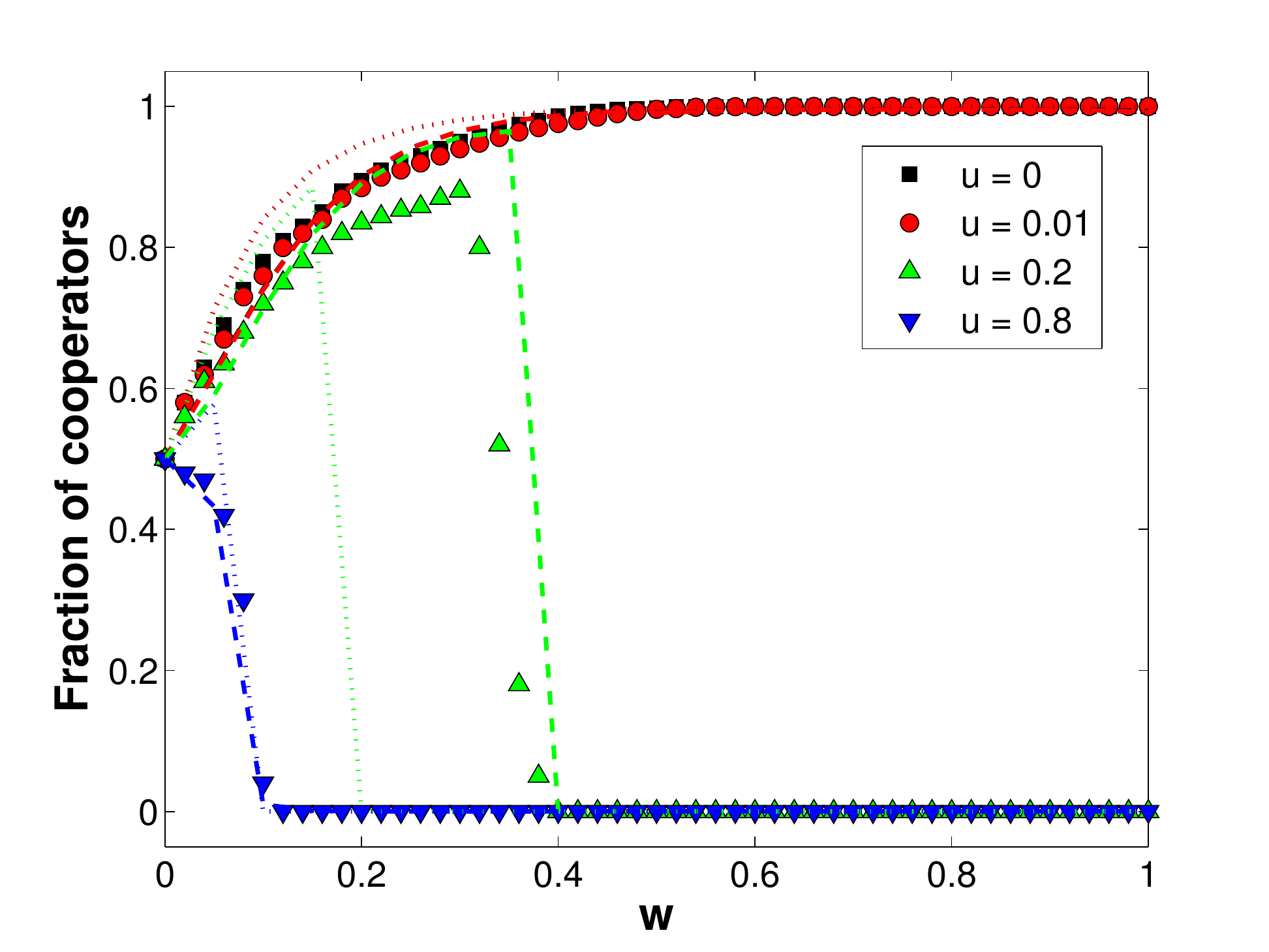}
\caption{(Left) Final fraction of cooperators versus cost-benefit ratio $u$ for different strategy updating probabilities $w$. Markers are averages from 1,000 simulations, dotted lines are pair approximation (PA) results, and dashed lines are from approximate master equations (AME). The PA results are different from those shown in \cite{fu2009partner}, but we have confirmed the accuracy of our results by personal communication with those authors. (Right) Final fraction of cooperators versus $w$ for different cost-benefit ratio $u$. Markers, dotted lines, and dashed lines again indicate simulations, PA, and AME, respectively. In both panels we observe that AME is typically more accurate than PA, especially so near the phase transition where the fraction of cooperators goes to zero.}
\label{fig:parameters}
\end{center}
\end{figure}

In Fig.~\ref{fig:parameters}, we further explore the final level of cooperation and assess the accuracies of the two semi-analytical approximations. In the left panel of Fig.~\ref{fig:parameters}, we consider $w = 0$, $w = 0.05$, $w = 0.1$, and $w = 0.5$, plotting the final fraction of nodes in state $C$ versus the cost-benefit ratio, $u$. There are three sets of outcomes here: simulations (markers), PA (dotted lines), and AME (dashed lines). For $w = 0$, there are no strategy updates and the final fraction of $C$ nodes is fixed at $1-\rho$. For $w > 0$, strategy updates come into play with larger values of the cost-benefit ratio $u$ driving stronger incentive for nodes to defect, decreasing the final fraction of cooperators. As seen in the figure, the cases $w = 0.1$ and $w = 0.5$ include higher levels of cooperation at small $u$ that decrease slightly as $u$ increases before suddenly dropping to zero near $u \doteq 0.2$ and $u \doteq 0.6$, respectively. This phenomenon is qualitatively captured by both PA and AME, though the quantitative description from AME is much more accurate, particularly in terms of these critical points. In contrast, the PA prediction of the fraction of cooperators drops to zero at much smaller values of $u$. Similarly, in the case $w = 0.05$, the level of cooperation in the simulation results decreases with increasing $u$, but without the sharp dropoff; in contrast to both the simulations and AME, the PA result drops to zero around $u \doteq 0.9$.

Similarly, in the right panel of Fig.~\ref{fig:parameters} we plot the final fraction of cooperators $C$ versus $w$ for different cost-benefit ratios, $u = 0$, $u = 0.01$, $u = 0.2$, and $u = 0.8$. For $w=0$, no nodes update their strategies and the final fraction of $C$ nodes is $1-\rho=0.5$. Larger $w$ increases both the number of strategy updates and their overall effect on the final state. In the cases $u = 0$ and $u = 0.01$, $u$ is so small that there is almost no incentive to defect, so in the strategy updating consideration $C$'s and $D$'s are, holding everything else equal, almost indistinguishable. However, in rewiring an edge the $C$ node drops its $D$ neighbor, reducing the number of partners for $D$, and over time the $D$ nodes experience lower payoffs because the numbers of their neighbors decrease. This decrease in degree results in lower total payoff, increasing the probability of a $D$ node changing to a $C$ strategy, leading to the dominance of $C$'s for small $u$. 

In contrast, for $u = 0.8$, the $C$ nodes become more disadvantaged with increased $w$ since the payoff of defecting is high, and the final fraction of cooperators decreases to zero around $w \doteq 0.1$. The case $u = 0.2$ is intermediate between the competing effects of these two extremes. The final fraction of cooperators increases for values of $w$ that are not too large due to the above-described rewiring effect. But sufficiently large strategy update rates (that is, fewer rewiring steps) combined with an apparently sufficient incentive to defect leads to final network states dominated by defectors. In the simulations, this transition for $u=0.2$ appears to occur near $w\doteq 0.4$. While the PA and AME results qualitatively describe all of these effects well, AME does a better job capturing this transition point. Taking these results together, we note that in general the increased frequency of breaking and rewiring $CD$ bonds (i.e., smaller $w$) can directly support cooperation, in agreement with \cite{szolnoki2008making, szolnoki2009resolving}, as observed here both in terms of the apparent phase transition and the behavior at large $u$ in the right panel of Fig.~\ref{fig:parameters}. But we also note that the results include situations where increasing this frequency (decreasing $w$) decreases the total level of cooperation, because of the complex interplay in the model dynamics.

\subsection{Network dynamics}

\begin{figure}
\begin{center}
\includegraphics[width=\textwidth]{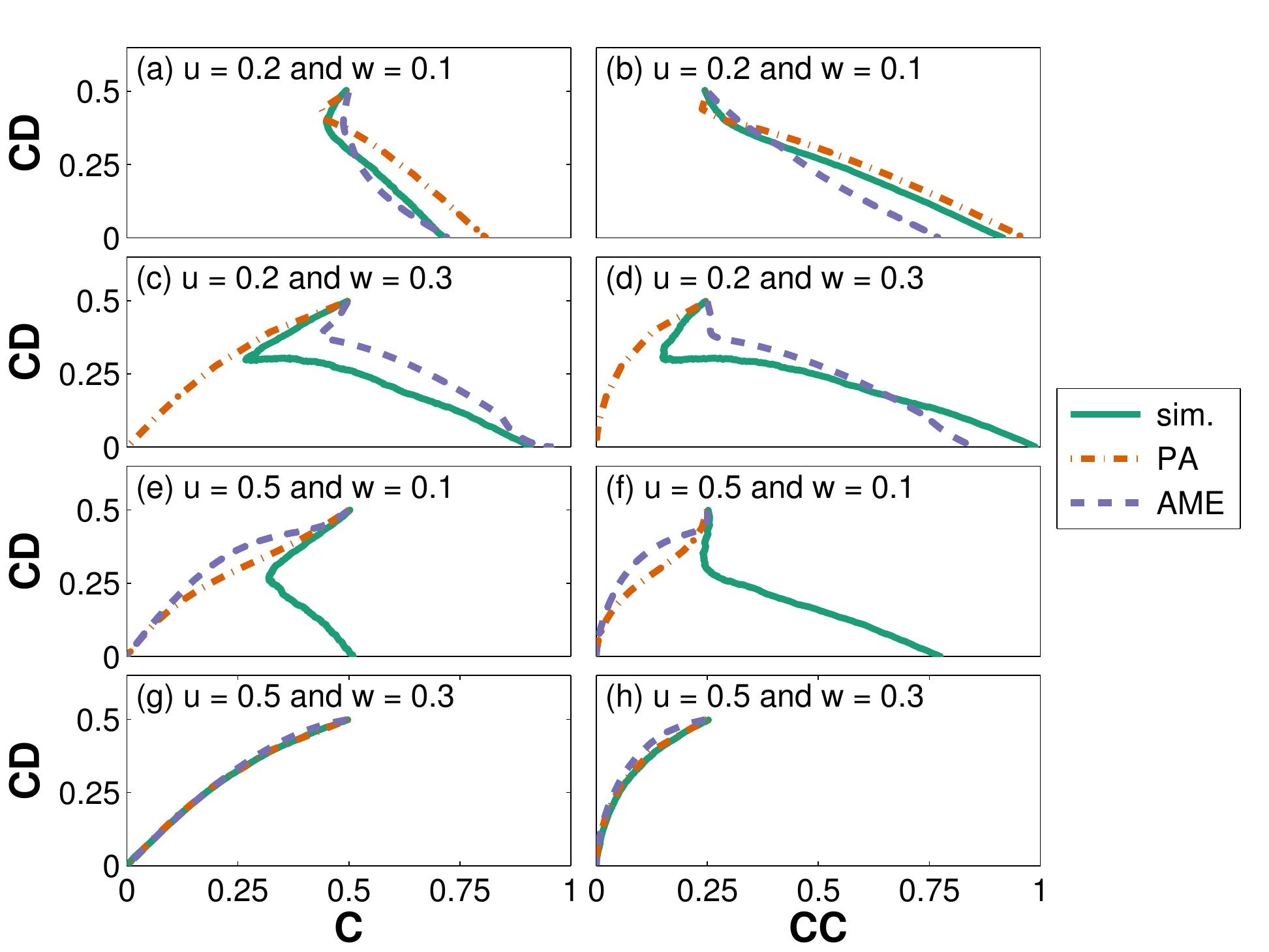}
\caption{Phase space visualizations of the temporal evolution of the fraction of edges that are $CD$ versus the fraction of nodes that are $C$ (left column) and versus the fraction of edges that are $CC$ (right column) for different values of the cost-benefit ratio $u$ and strategy updating probability $w$ (rows): (a\&b) $u = 0.2$, $w = 0.1$, (c\&d) $u = 0.2$, $w = 0.3$, (e\&f) $u = 0.5$, $w = 0.1$, and (g\&h) $u = 0.5$, $w = 0.3$.  Green lines are the averages of 50 simulations results, red dash-dot lines are the pair approximation (PA), and blue dashed lines are the semi-analytical results of approximate master equations (AME). All networks start at $C=0.5$, $CC=0.25$, $CD=0.5$. As the system evolves, on average the number of $CD$ decreases over time until none of these discordant edges remain ($CD=0$). Sufficiently far from the phase transition (as in the top and bottom rows), both approximations are reasonably good. Near the phase transition there are ranges of parameters where only $AME$ provides a good description (such as in the second row, where $PA$ incorrectly predicts the position relative to the phase transition; compare with the green markers and lines in the right panel of Fig.~\ref{fig:parameters}). Also near the phase transition, there are other parameters where neither approximation captures the qualitative behavior (as in the third row, with $u=0.5$ and $w=0.1$; compare with the location of the observed and predicted phase transition in Fig.~\ref{fig:phasediagram} and with the green markers and lines in the left panel of Fig.~\ref{fig:parameters}).}
\label{fig:phasespace}
\end{center}
\end{figure}

Having explored the populations in the final states, we consider how the networks evolve to those states. Specifically, we investigate the evolution of five fundamental quantities of the networks: the fractions of nodes in states $C$ and $D$, and the fractions of edges that are $CC$, $CD$, and $DD$. We recall that the total numbers of nodes and edges are constant during the dynamics, requiring both $C+D$ fractions and $CC+CD+DD$ fractions to be equal to $1$. We visualize the average trajectories through this phase space for a collection of different $(u,w)$ parameters in Fig.~\ref{fig:phasespace}, starting from an initial fraction of defectors $\rho=0.5$ distributed across an Erd\H{o}s-R\'{e}nyi $G(N, M)$ random graph, comparing simulations with PA and AME predictions. As such, the initial fraction of $CC$, $DD$, and $CD$ edges is $0.25$, $0.25$, and $0.5$, respectively; that is, all networks start at $C=0.5$, $CC=0.25$, $CD=0.5$. As the networks evolve, these fractions change, tracing out trajectories in the $(C,CD)$ and $(CC,CD)$ coordinates in the figure. We perform 50 simulations at each set of parameter values and compute the averages across simulations at each time step. The networks evolve until there are no discordant edges (i.e.\ $CD = 0$). 

For comparison, we include both PA and AME in the figure. While some features of the trajectories are better captured than others, both methods in general do reasonably well qualitatively, with better overall accuracy for AME, as seen in the figure. We note that both approximations are qualitatively incorrect for the case $u=0.5$, $w=0.1$. Recalling the simulation and AME results in the Fig.~\ref{fig:phasediagram} phase diagram, we note that this parameter set is within the relatively narrow range between the observed location of the phase transition and the AME prediction.

\subsection{Degree distributions}

\begin{figure}
\begin{center}
\includegraphics[width=0.9\textwidth]{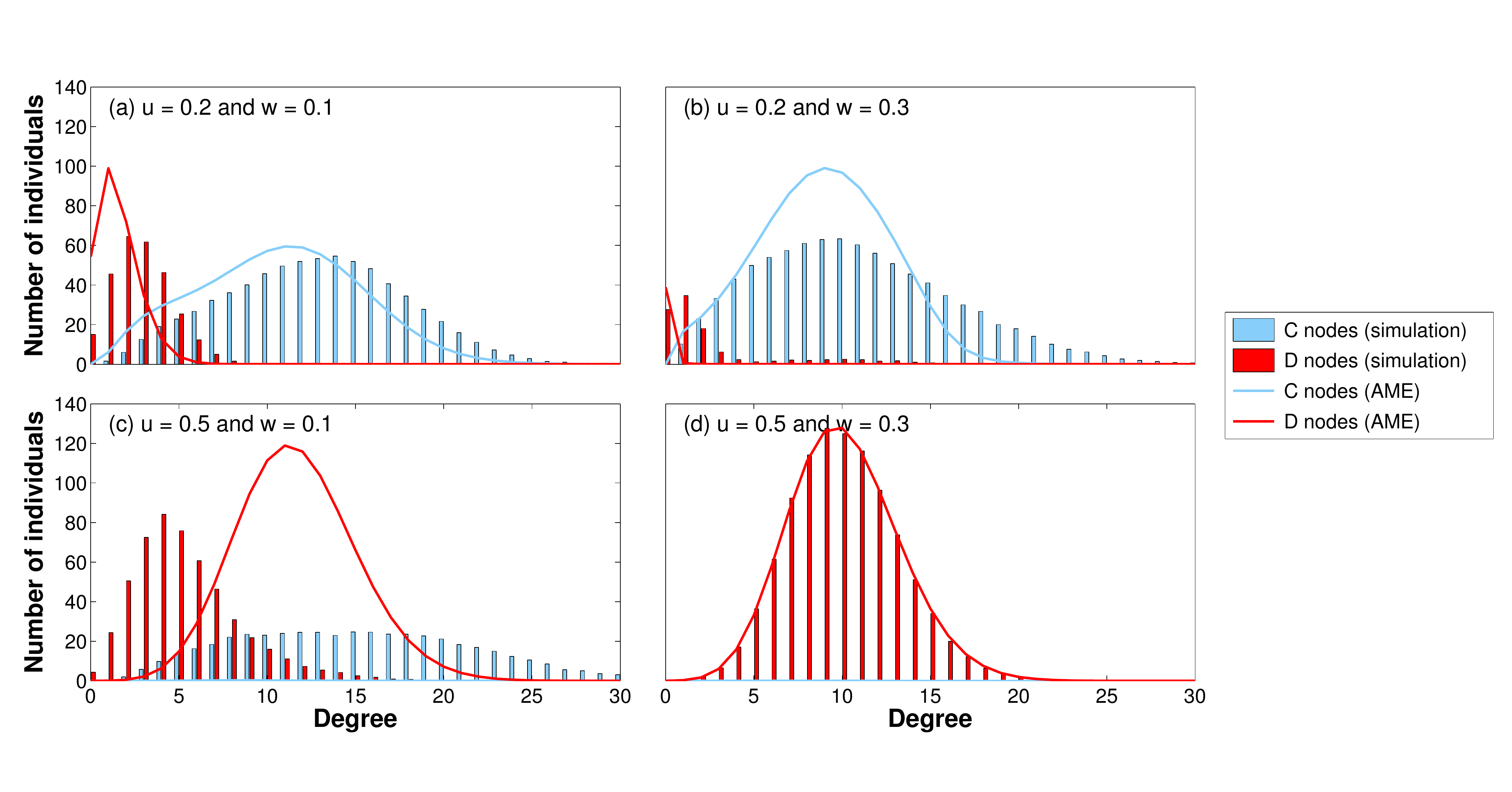}
\caption{Degree distributions in the stationary states for different values of the cost-benefit ratio $u$ and strategy updating probability $w$: (a) $u = 0.2$, $w = 0.1$; (b) $u = 0.2$, $w = 0.3$; (c) $u = 0.5$, $w = 0.1$; and (d) $u = 0.5$, $w = 0.3$. Bars indicate averages from 50 simulations, with colors distinguishing the degree distributions of cooperator (blue) and defector (red) nodes. Blue and red lines indicate the prediction from the semi-analytical approximate master equations (AME). We note the qualitatively good agreement of the AME prediction with simulations, except in case (c) where AME predicts extinction of the cooperator nodes, as we have previously seen (see Figure \ref{fig:parameters}e\&f).}   \label{fig:degreedist}
\end{center}
\end{figure}

The AME method utilizes more information and has higher computational cost than PA, but as we have seen it generally provides a better approximation for the current model. Moreover, and unlike PA, AME explicitly includes information about degree distributions. In Fig.~\ref{fig:degreedist}, we plot the final state degree distributions of $C$ and $D$ nodes, comparing the simulations and AME predictions. In each case, since we initialize with random strategies on the Erd\H{o}s-R\'{e}nyi $G(N, M)$ random graph model, the initial degree distributions are approximately Poisson (for large $N$). We examine degree distributions in the final states, revisiting the same parameters considered previously: $u = 0.2, w = 0.1$; $u = 0.2, w = 0.3$; $u = 0.5, w = 0.1$; and $u = 0.5, w = 0.3$. In the case $u = 0.5, w = 0.3$, it is perhaps not surprisingly that AME gives an excellent prediction of the final degree distribution, since we observed excellent prediction of the trajectory of the dynamics in this case above (Fig.~\ref{fig:phasespace}) . In the final state of this case, $D$ nodes dominate the whole network, with no $C$ nodes left. 

In the other three cases, although the AME predictions are not as accurate, we can still see the AME provides a qualitative picture of the final degree distribution. Once again, the worst case of the four parameter sets visualized here is the $u=0.5$, $w=0.1$, where the AME prediction is on the wrong side of the phase transition.

\subsection{The effect of the initial fraction of defectors, $\rho$}

\begin{figure}
\begin{center}
\includegraphics[trim={8.8cm 0cm 11cm 0cm},clip,width=0.47\textwidth]{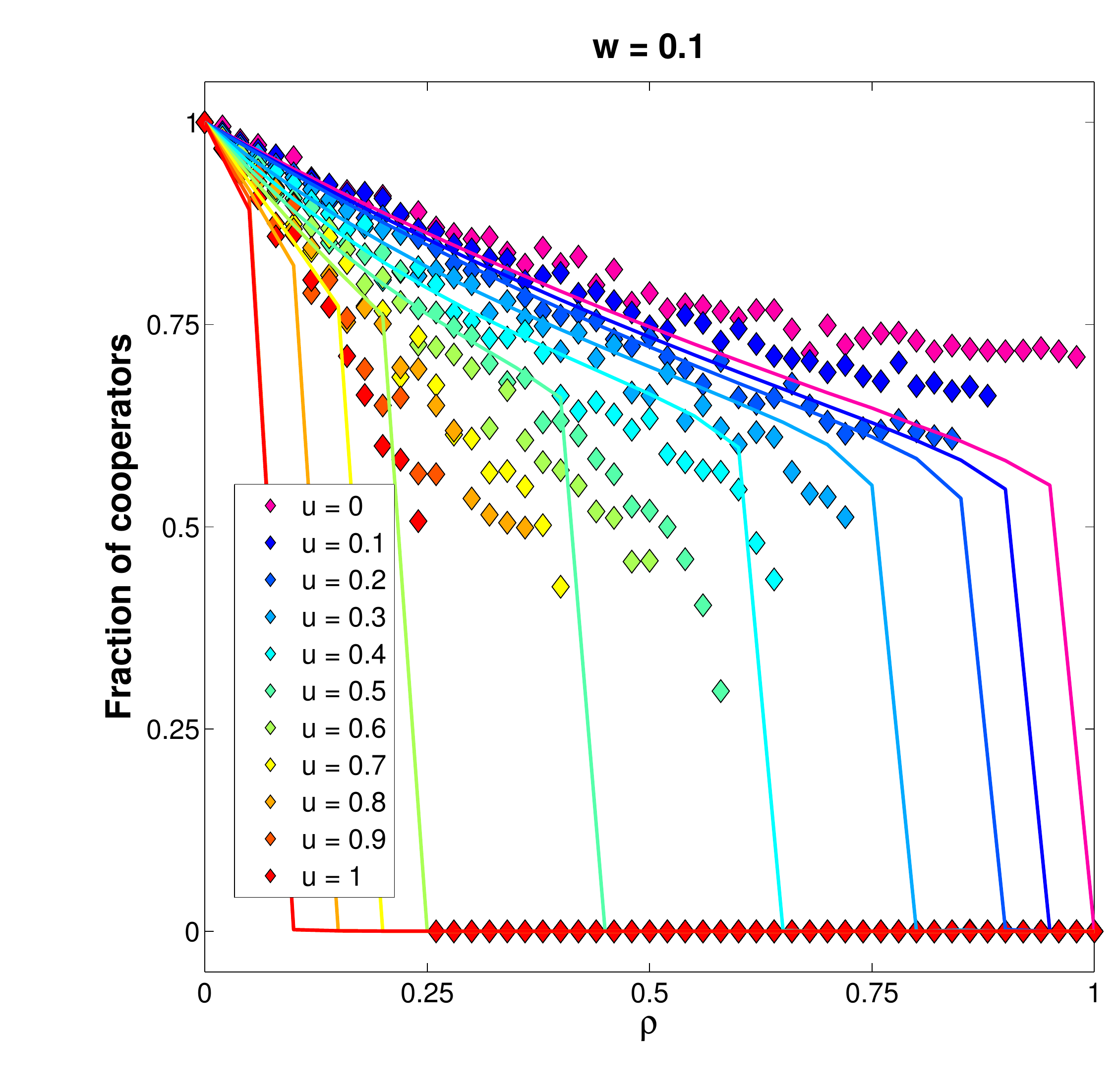}
\quad
\includegraphics[trim={8.8cm 0cm 11cm 0cm},clip,width=0.47\textwidth]{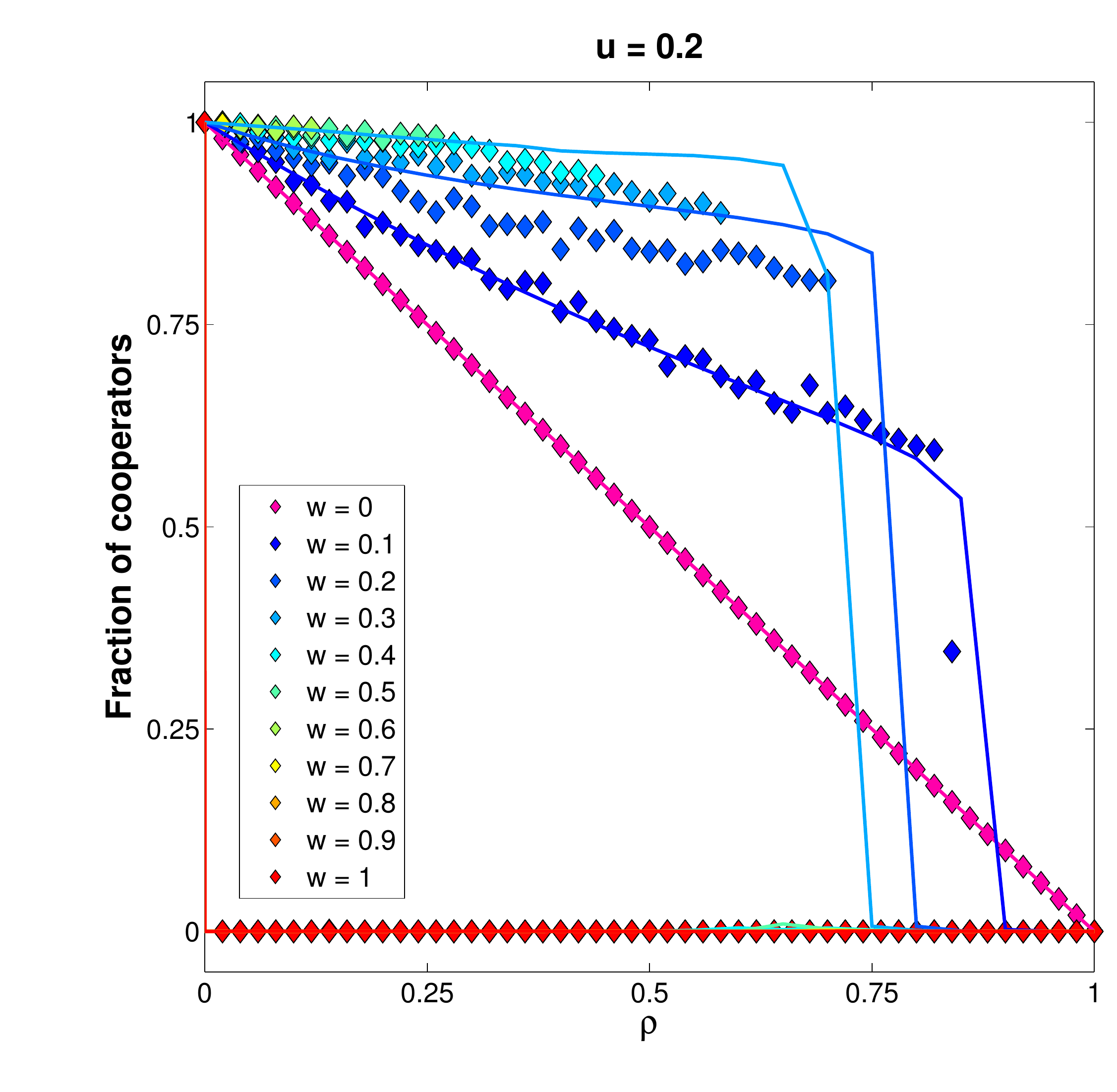}
\caption{(Left) Fraction of cooperators remaining in the stationary state versus the initial fraction of defectors $\rho$, for strategy updating probability $w = 0.1$ and various cost-benefit ratios $u$: 0, 0.1, ..., 1. (Right) Fraction of cooperators versus $\rho$ for $u = 0.2$ and various strategy updating probabilities $w$: 0, 0.1, ..., 1.. In both figures, markers are the averages from 50 simulations and the lines are the semi-analytical predictions from approximate master equations (AME).}
\label{fig:initfraction}
\end{center}
\end{figure}

In Fig.~\ref{fig:initfraction} we explore the effect of the initial fraction of defectors, $\rho$. In this figure, we fix the parameter to be $w = 0.1$ in the left panel and $u = 0.2$ in the right panel, changing the other parameter ($u$ and $w$, respectively) to view the final fraction of cooperators against $\rho$ in stationary states, comparing simulation results (markers) with our AME approximation (lines). By construction, all results in these panels should connect the upper left and lower right corners because when $\rho = 0$ (respectively, $\rho = 1$) there are no $D$ ($C$) nodes, no discordant edges, and no dynamics. 

In the left panel, for the different $u$ curves, the final fraction of cooperators first gradually decreases and then drops to zero suddenly at different values of $\rho$. AME predicts this qualitative behavior for different levels of $u$. Moreover, fixing $w = 0.1$, smaller $u$ results in higher cooperation, since the incentive of defecting is less when $u$ is small. 

In the right panel, fixing $u = 0.2$, the diagonal line is the case $w = 0$, where there are no strategy updates and the cooperation level remains the same during the evolution. For small and moderate $w$, the curves decrease with increasing $\rho$ (though more quickly for smaller $w$), followed by a sharp drop to zero at decreasing values of $\rho$ for increased $w$. When $w$ is large (greater rates of strategy updates), the final cooperative level drops to zero for all non-zero $\rho$. As we show in this panel, the AME curves suddenly drop to zero when $w \geq 0.4$. 

\section{A model variant with $DD$ rewiring}
\label{sec:variation}

In this Section, we study a slightly different variant of the partner switching evolution game model, also introduced by \cite{fu2009partner}, where $DD$ edges can also rewire as defector nodes seek out new partners that they might better exploit. We are particularly interested in how this additional rewiring affects the final cooperation level. In this variant, each time step starts by uniformly at random selecting from the union of $CD$ and $DD$ edges. (Recall that only $CD$ edges were selected in the original model above.) If a $CD$ edge is picked, the dynamics proceed as in the original model: with probability $w$, the strategy update process happens; otherwise (that is, with probability $1-w$) the $C$ node along this edge unilaterally drops the partnership with its $D$ neighbor and rewires to another node. In contrast, if a $DD$ edge is picked, then with probability $w$ nothing happens (there is no strategy update); otherwise (that is, with probability $1-w$) one of the two $D$ ends, selected with equal probability, drops the connection and rewires uniformly at random to another node in the network to whom it is not already connected. We note that the total numbers of nodes, $N$, and edges, $M$, remain constant, as in the original model.

\begin{figure}
\begin{center}
\includegraphics[trim={0cm 4cm 0cm 4cm},clip,width=0.45\textwidth]{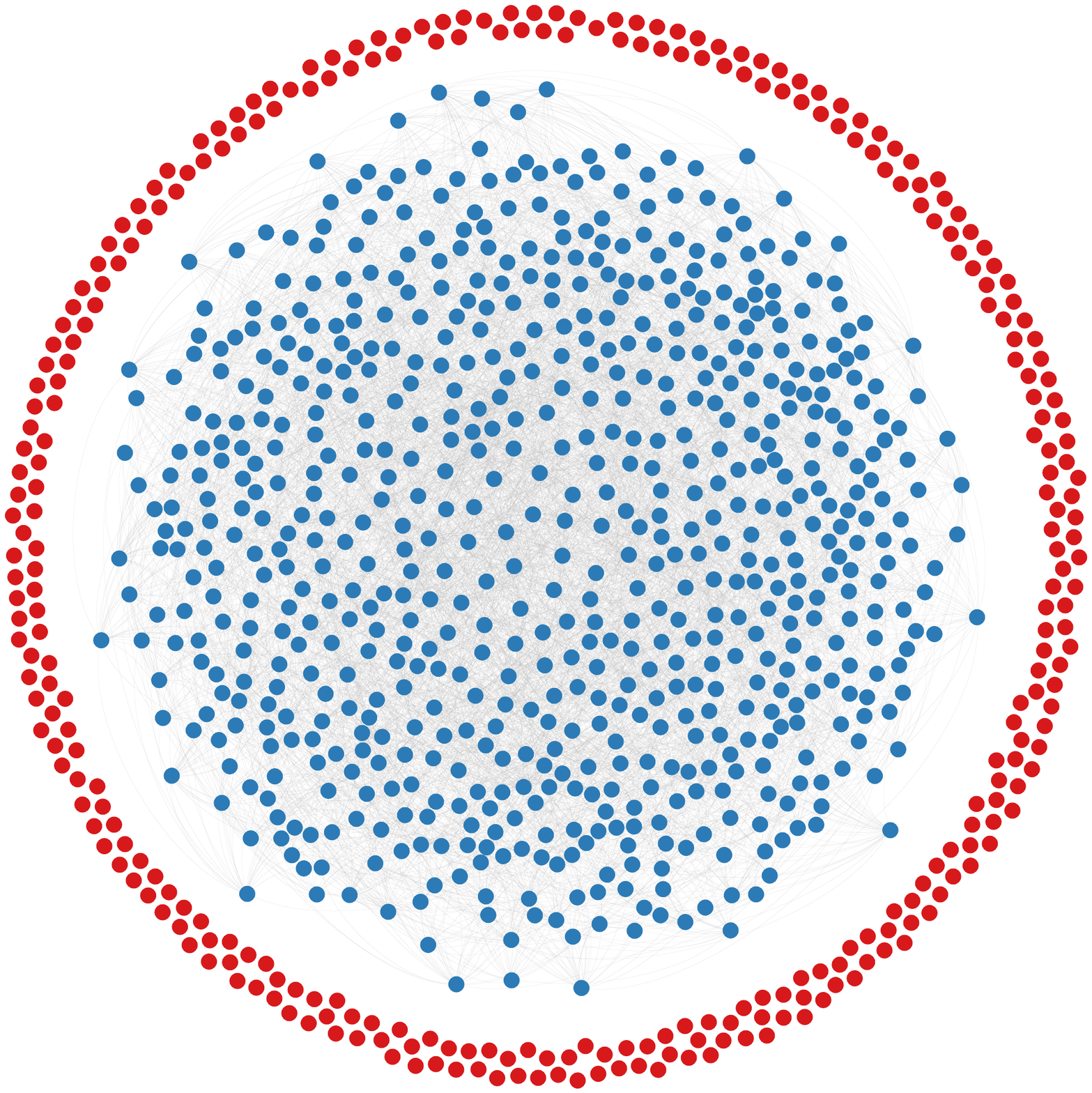}
\qquad
\includegraphics[trim={0cm 5.4cm 0cm 5.5cm},clip,width=0.47\textwidth]{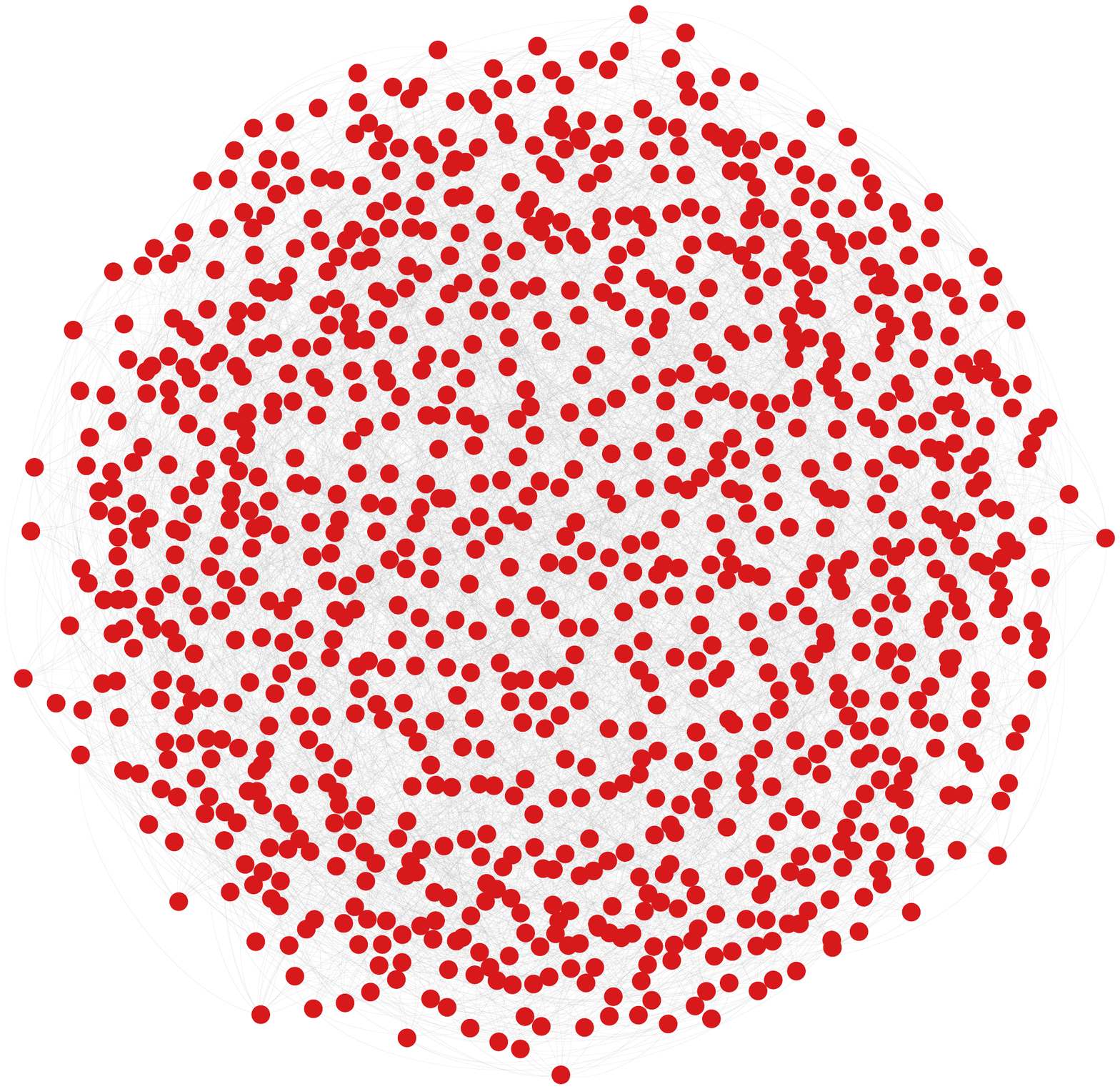}
\caption{Visualization of stationary states of the model variant with $DD$ rewiring, for $N = 1,000$ nodes, $M = 5,000$ edges, and initial fraction of defectors $\rho = 0.5$, for different parameters. Colors correspond to the two node states: cooperating (blue) and defecting (red). (Left) For cost-benefit ratio $u = 0.5$ and strategy updating probability $w = 0.1$, $D$ nodes remain in the system but they have each been isolated with zero degree. (Right) For $u = 1$ and $w = 0.5$, the system evolves until only $D$ nodes exist in the network and all the edges are $DD$ edges. These visualizations were created using the Yifan Hu layout in Gephi.}
\label{fig:visualization2}
\end{center}
\end{figure}

This variant model stops evolving when there are no $CD$ or $DD$ edges remaining in the system; that is, all edges in a final, frozen state are $CC$. However, this does not require that all $D$ nodes are removed from the system, only that they are each left isolated with no connections. Another possibility, in which the model variant as stated above never reaches a stopping condition, is that the system evolves to a statistically stationary state where all $C$ nodes have been removed so that all edges are $DD$ edges and the continued rewiring of these $DD$ edges simply serves to repeatedly randomize the network of defector nodes. Since in this case there are no $C$ nodes left to emulate in a strategy update nor to connect to in a rewiring, we do not consider the dynamics further beyond this event. Both of these cases are visualized in Fig.~\ref{fig:visualization2}. In the left panel, the $C$ nodes own all of the edges, having isolated the exploitation efforts of the $D$ nodes. In the right panel, only $D$ nodes remain and all edges are $DD$.

\subsection{Semi-analytical methods of approximation}
As before, we study this model variant with a combination of simulations and approximate analytic models, comparing the pair approximation (PA) and approximate master equation (AME) approaches with simulation results. 

Following the notation of the original model in Sec.~\ref{sec:samethods}, the PA equations for this variant become.  
\begin{equation}
\begin{split}
\frac{dN_{C}}{dt} = & w\cdot N_{CD}\cdot \tanh \bigg[ \frac{\alpha}{2}( \overline{\pi}_{C} - \overline{\pi}_{D}) \bigg] \\
\frac{dN_{CC}}{dt} = & w\cdot \bigg( N_{CD}\phi_{C \rightarrow D} - 2N_{CD}\frac{N_{CC}}{N_{C}}\phi_{D \rightarrow C} + N_{CD}\frac{N_{CD}}{N_{D}}\phi_{C \rightarrow D} \bigg) \\
& + (1-w) \cdot \frac{N_{C}}{N}N_{CD}  \\
\frac{dN_{DD}}{dt} = & w \cdot \bigg( N_{CD}\phi_{D \rightarrow C} - 2N_{CD}\frac{N_{DD}}{N_{D}}\phi_{C \rightarrow D} + N_{CD}\frac{N_{CD}}{N_{C}}\phi_{D \rightarrow C} \bigg) \\
& - (1-w) \cdot \frac{N_{C}}{N}N_{DD}  
\end{split}
\label{eq:PADD}
\end{equation}
where we note the only difference between equation (\ref{eq:PADD}) and equation (\ref{eq:PA}) is the appearance of the last term in the $dN_{DD}/dt$ equation. This term captures the possibility in this variant model for a $D$ node to dismiss its defective partner and rewire to a $C$ node, thus decreasing the $DD$ count.

Similarly, we modify the AME equations to account for the rewiring of $DD$ edges, yielding
\begin{equation}
\begin{split}
\frac{dC_{k,l}}{dt} & = w \bigg\{ \phi^{D}_{k,l} \cdot (k-l)D_{k,l} - \phi^{C}_{k,l} \cdot l C_{k,l}\\
& + \phi_{CD \leftarrow CC} \cdot \gamma^{C}(l+1) C_{k,l+1} - \phi_{CD \leftarrow CC} \cdot \gamma^{C} l C_{k,l}\\
& + \phi_{CC \leftarrow CD} \cdot \beta^{C}(k-l+1) C_{k,l-1} - \phi_{CC \leftarrow CD} \cdot \beta^{C}(k-l) C_{k,l} \bigg \}\\
& + (1-w) \bigg\{ \frac{N_{C}}{N} \big[(l+1) C_{k,l+1} - l C_{k,l}\big]
 + \frac{N_{CD}}{N_{}} \big[C_{k-1,l} - C_{k,l}\big]\\
& + \frac{N_{DD}}{N} \big[ C_{k-1,l-1} - C_{k,l} \big] \bigg \}
\end{split}
\label{eq:AME-C-DD}
\end{equation}
\begin{equation}
\begin{split}
\frac{dD_{k,l}}{dt} & = w \bigg\{ -\phi^{D}_{k,l} \cdot (k-l) D_{k,l} + \phi^{C}_{k,l} \cdot l C_{k,l}\\
& + \phi_{DD \leftarrow DC} \cdot \gamma^{D}(l+1) D_{k,l+1} - \phi_{DD \leftarrow DC} \cdot \gamma^{D} l D_{k,l}\\
& + \phi_{DC \leftarrow DD} \cdot \beta^{D}(k-l+1) D_{k,l-1} - \phi_{DC \leftarrow DD} \cdot \beta^{D} (k-l) D_{k,l} \bigg \}\\
& + (1-w) \bigg\{ \big[(k-l+1) D_{k+1,l} - (k-l) D_{k,l}\big] + \frac{N_{CD}}{N_{}} \big[D_{k-1,l} - D_{k,l}\big] \\
& + \frac{N_{C}}{2N} \big[(l+1)D_{k,l+1} - l D_{k,l}\big]  + \frac{N_{DD}}{N_{}} \big[D_{k-1,l-1} - D_{k,l}\big] \\
& + \frac{1}{2}\big[(l+1)D_{k+1,l+1} - l D_{k,l} \big]  \bigg \}
\end{split}
\label{eq:AME-D-DD}
\end{equation}
Again, the additional terms here compared to equations (\ref{eq:AME-C}) and (\ref{eq:AME-D}) account for the rewiring of $DD$ edges, accounting for the rates of $D$ nodes abandoning $D$ neighbors and rewiring to $C$ nodes, $D$ nodes passively gaining new connections through this rewiring, and $D$ nodes being abandoned by their $D$ neighbors. Note the $1/2$ factors in the last two lines of the $D_{k,l}$ equation, accounting for an individual node on a $DD$ keeping and losing the edge.

\subsection{Final level of cooperation}

\begin{figure}
\begin{center}
\hspace*{-0.25in}
\includegraphics[width=0.5\textwidth]{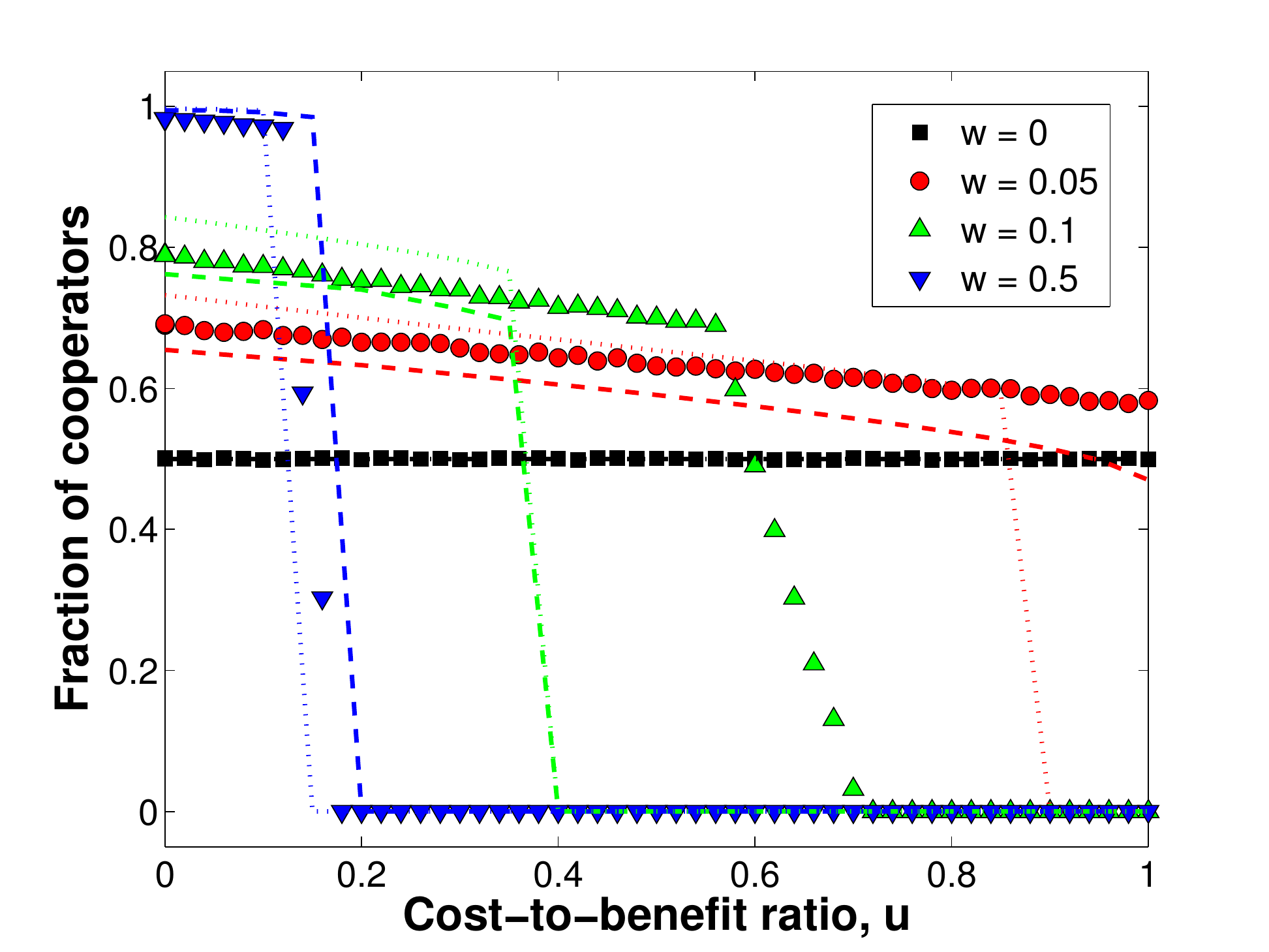}
\includegraphics[width=0.5\textwidth]{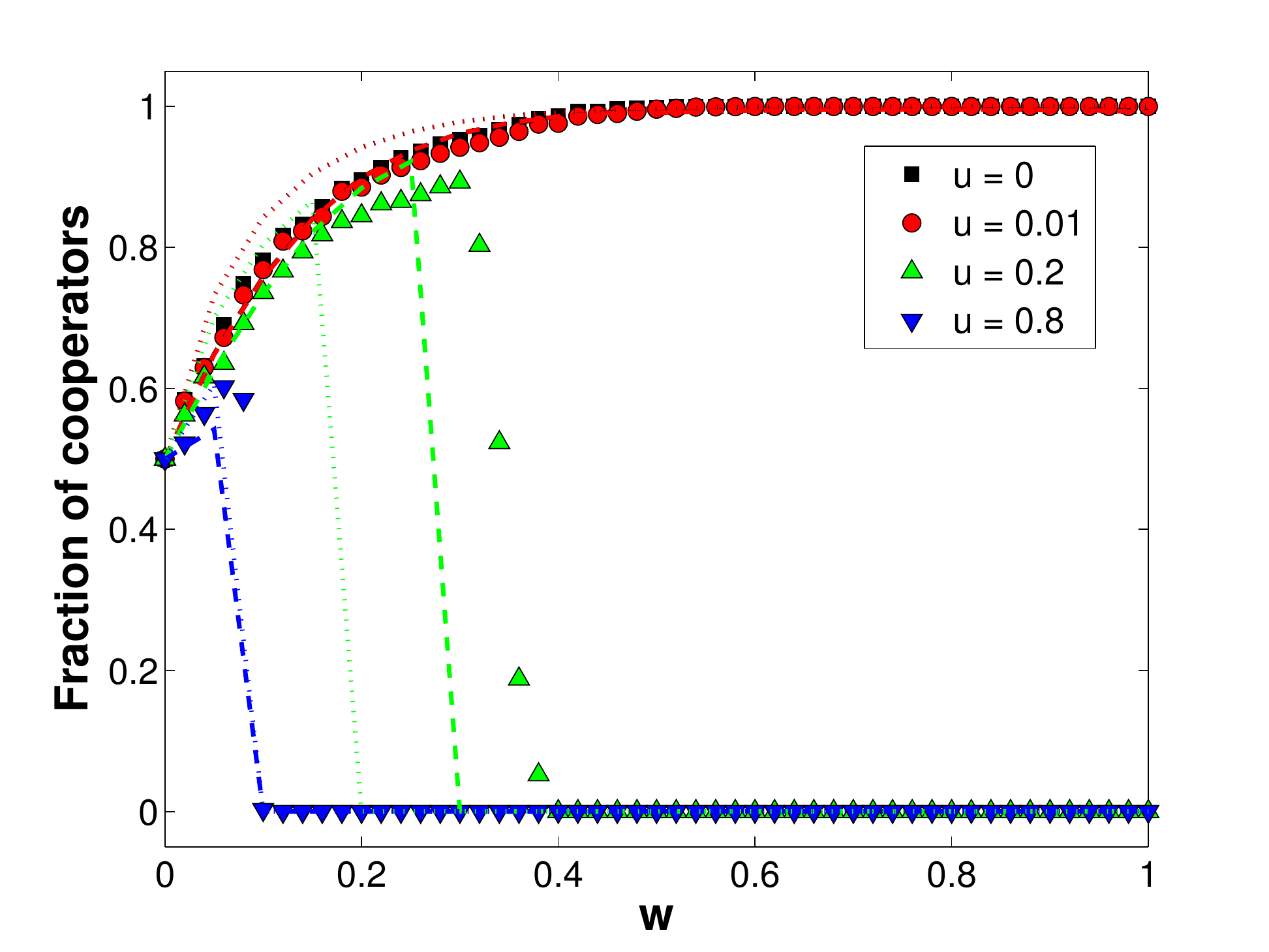}
\hspace*{-0.25in}
\caption{(Left) Fraction of cooperators remaining in the stationary state of the model variant with $DD$ rewiring versus cost-benefit ratio $u$, for different strategy updating probabilities $w$. Markers are averages from 1,000 simulations, dotted lines are the semi-analytical results from pair approximation (PA), and the dashed lines are the semi-analytical results from approximate master equations (AME). (Right) Fraction of cooperators remaining in the stationary state versus $w$ for different $u$ values, with markers and lines as in the left panel.}
\label{fig:initfraction2}
\end{center}
\end{figure}

As before, we compare the final fraction of cooperators in the simulations versus the PA and AME predictions. In the left panel of Fig.~\ref{fig:initfraction2}, we consider $w = 0$, $w = 0.05$, $w = 0.1$, and $w = 0.5$, plotting the fraction of $C$ nodes in the final states versus the cost-benefit ratio, $u$. In the right panel, we fix values of $u$ and vary $w$. Again, markers indicate simulation results, dotted lines are from the PA equations, and dashed lines are from AME.

Comparing Fig.~\ref{fig:initfraction2} for this variant with the results in Fig.~\ref{fig:parameters}, the inclusion of $DD$ rewiring here leads to slightly larger final fractions of $C$ nodes for small $w>0$. In particular, whereas an existing $DD$ link in the original model can only be changed if the state of one of the two $D$ nodes changes in a strategy update, the $DD$ rewiring in this variant increases the numbers of $CD$ edges and thus leaves open the possibility of $C$ nodes becoming more favorable because of their yet higher degree. However, for larger values of $w$, the more rapid rate of strategy updating leads to defectors dominating the system before the cooperators can gain any advantage from increased degrees.

\begin{figure}
\begin{center}
\includegraphics[width=0.475\textwidth]{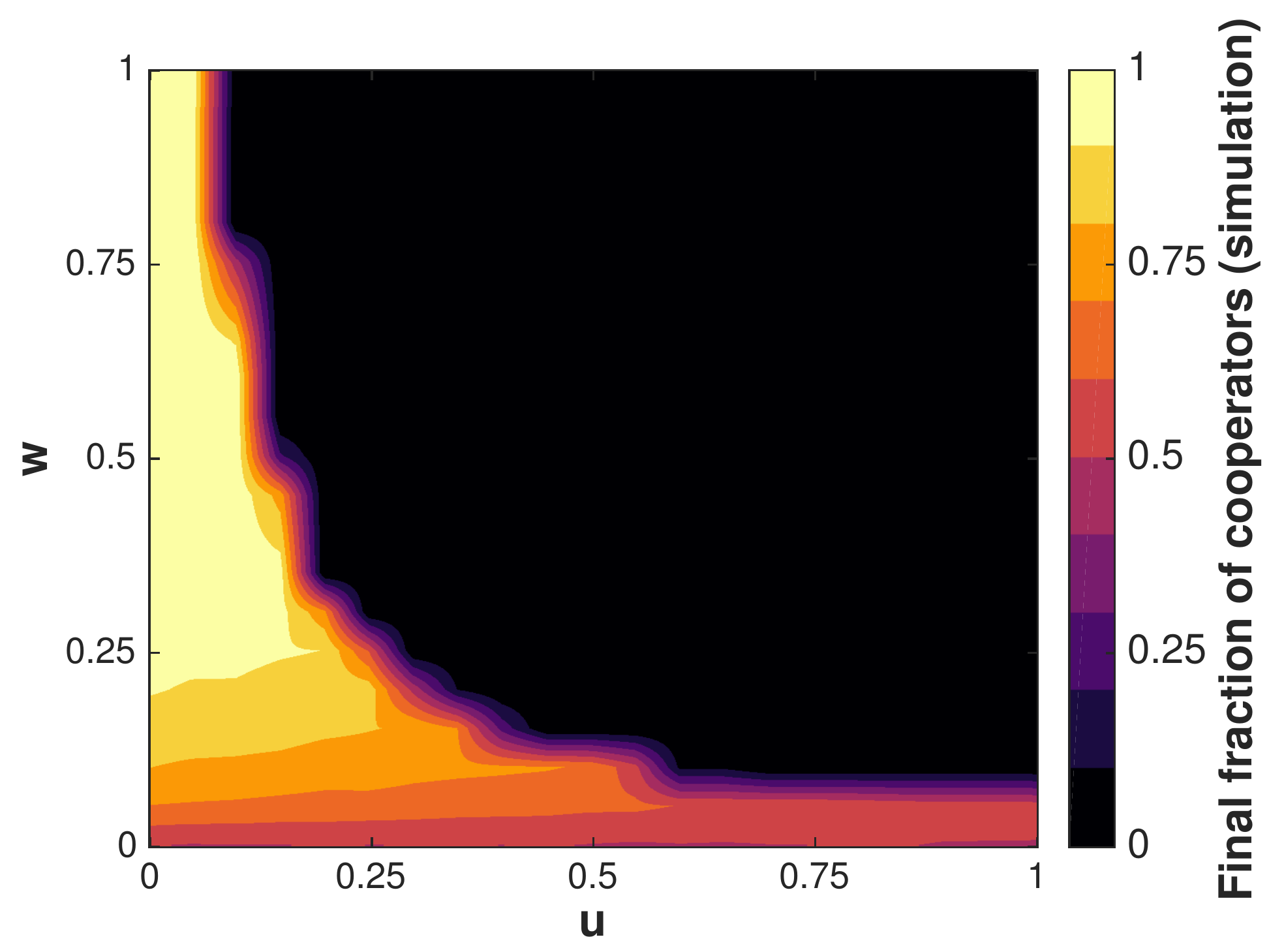}\quad
\includegraphics[width=0.475\textwidth]{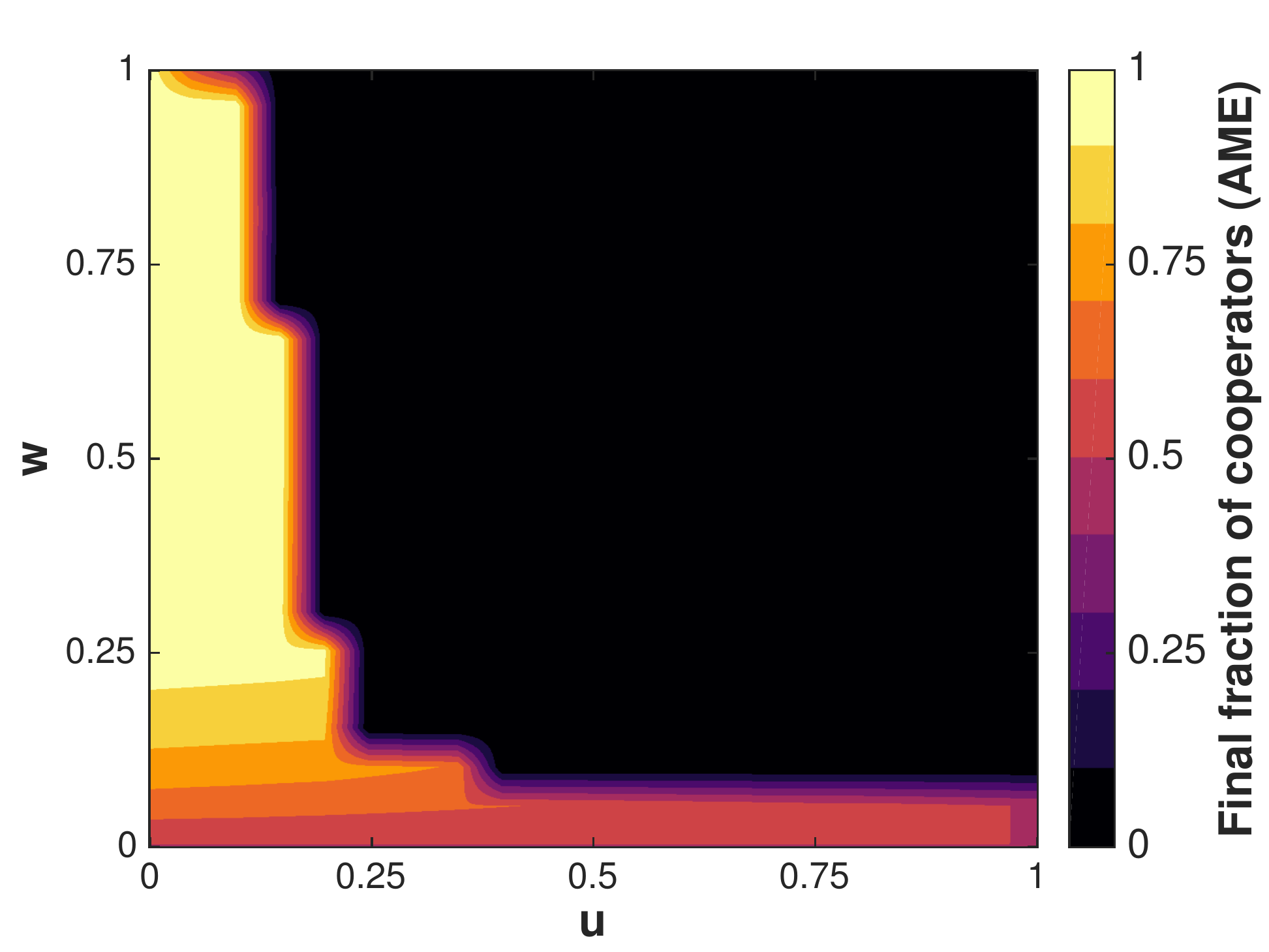}

\includegraphics[width=0.475\textwidth]{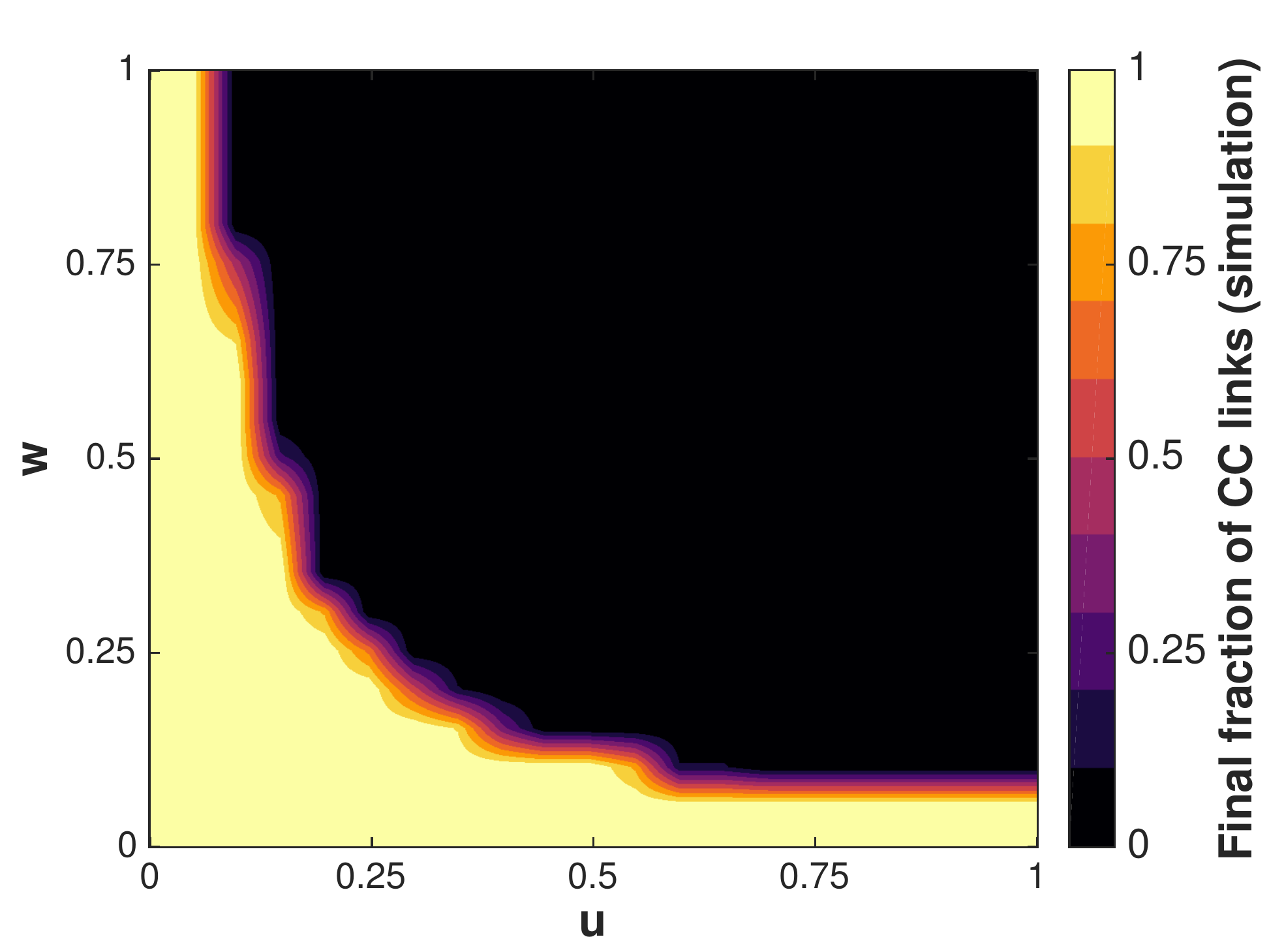}\quad
\includegraphics[width=0.475\textwidth]{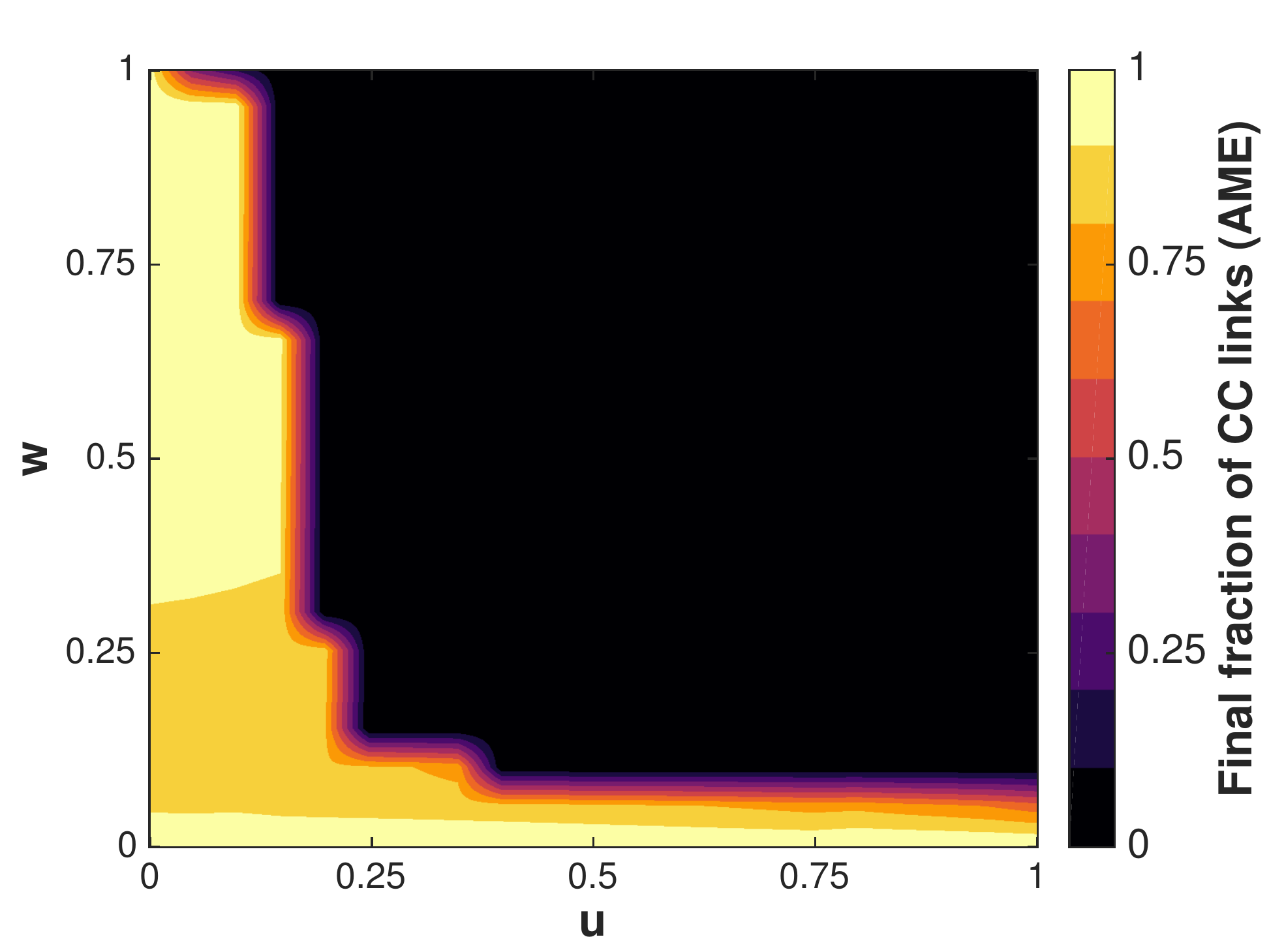}

\caption{Results from simulations (left column) and Approximate Master Equations (AME) (right column) of the final fractions of $C$ nodes (top row) and $CC$ links (bottom row) for different combinations of the cost-benefit ratio $u$ and strategy updating probability $w$ in the model variant with $DD$ rewiring. The initial fraction of defectors is $\rho=0.5$. For both the $u$ and $w$ axes, we use steps of 0.05 and plot the results from stationary states. Simulation results here are averaged over 50 realizations at each parameter set. These visualizations were generated from results on a regular grid through bilinear interpolation, leading to some clearly apparent grid artifacts. As in the original model without $DD$ rewiring (Fig.~\ref{fig:phasediagram}), while some discrepancies between simulation and AME results are clearly present, we note in particular that the position of the phase transition in the $(u,w)$ parameter space is well approximated by the AME system.}
\label{fig:phasediagram2}
\end{center}
\end{figure}

For further comparison, we consider the final fractions of $C$ nodes and $CC$ edges for different combinations of $u$ and $w$ in Fig.~\ref{fig:phasediagram2}, as compared with the similar plots in Fig.~\ref{fig:phasediagram}. While many of the general features remain the same in the presence of this DD-rewiring variant, a notable difference is in the final fraction of $CC$ edges for small $w$. In particular, at $w=0$ (no strategy updates), the original model results visualized in Fig.~\ref{fig:phasediagram} limit to a final $CC$ level of 0.75, since the initial $DD$ edges (at 0.25 for $\rho=0.5$) cannot ever rewire under these settings. But these edges of course do rewire in the $DD$-rewiring variant, leading $CC$ to limit to 1 as $w\to 0$.

\section{Conclusion}
We have investigated a node-based Prisoner's dilemma game played on a network coevolving with player strategy updates. Our study includes two model variants for rewiring: one where only $CD$ edges can rewire, by $C$ nodes dropping a link to a $D$ neighbor in favor of a new partner; and a variant where $DD$ edges can also rewire. We explore the parameter space to investigate the competing effects of strategy updates and partner switching, as well as the initial levels of cooperation versus defection. We compare our simulations to the existing pair approximation (PA) developed in \cite{fu2009partner}, and we develop an approximation using approximate master equations (AME) to more accurately capture the transitions between the properties of the final states of these models. We also use the AME method to estimate the final-state degree distributions for different parameters.

We are particularly interested in the features that determine the final level of cooperation and overall utility in the network. Revisiting Fig.~\ref{fig:parameters}, and the corresponding results for the model variant in Fig.~\ref{fig:initfraction2}, it is clear that the most effective way to increase the final fraction of cooperators is to directly reduce the cost-benefit ratio, $u$, so that players do not have as much incentive to defect. Alternatively, if the strategy update rate $w$ is small enough, the effects of rewiring give greater advantages to $C$ nodes as they accumulate larger numbers of playing partners, which then results in larger numbers of cooperators through the strategy updates. Of course, one can also directly decrease the initial fraction of defectors.

The question of how to maximize the total utility or payoff relates to the types of edges.  In the partner-switching evolutionary game model we study here, the total number of nodes and edges remain fixed, no matter how the networks and node states change. Utility comes from every edge in the network, with each $CC$, $CD$, and $DD$ edge contributing (from the two nodes) $2$, $1+u$, and $2u$, respectively, where $u \in (0,1)$. With no $CD$ edges in the final state, maximizing the final overall utility reduces to maximizing the number of $CC$ links. In particular, comparing the right panels of Figs.~\ref{fig:phasediagram} and \ref{fig:phasediagram2} for the two model variants, whether or not $DD$ edges can rewire significantly affects the $CC$ count (and thus the total system utility) for small strategy update rates.

\section*{Acknowledgments}
Research reported in this publication was supported by the Eunice Kennedy Shriver National Institute of Child Health and Human Development of the National Institutes of Health under Award Number R01HD075712. The content is solely the responsibility of the authors and does not necessarily represent the official views of the National Institutes of Health. We are grateful to the reviewers for identifying multiple important references cited here.



\bibliographystyle{comnet}
\bibliography{games}
%


\end{document}